\documentclass{JHEP3}
\usepackage{amsmath,amsfonts,amsthm,graphicx}
\usepackage{ulem}

\newcommand{\bal}{\begin{align}}
\newcommand{\eal}{\end{align}}
\newcommand{\beq}{\begin{equation}}
\newcommand{\eeq}{\end{equation}}
\newcommand\beqa{\begin{eqnarray}}
\newcommand\eeqa{\end{eqnarray}}
\newcommand\bea{\begin{array}}
\newcommand\eea{\end{array}}
\newcommand\comment[1]{{}}

\newcommand{\eq}[1]{(\ref{#1})}

\renewcommand{\sl}{\frak{sl}}

    \newcommand{\nn}{\nonumber}
    
    \newcommand{\COMMENT}[1]{}
    
    \newcommand{\neqa}{\nonumber\end{eqnarray}}
    \newcommand{\la}[1]{\label{#1}}

\def\a{{\alpha}}

\def\[{\left[}
\def\]{\right]}

\def\D{\Delta}
\def\l{\lambda}

\def\a{\alpha}

\def\[{\left[}
\def\]{\right]}

\def\<{\langle}
\def\>{\rangle}

\def\i2{\frac{i}{2}}

\def\be{\begin{eqnarray}}
    \def\ee{\end{eqnarray}}

    \def\CC{{\cal C}}

    \def\CI{{\cal I}}
    \def\CJ{{\cal J}}

    \def\CM{{\cal M}}
    
    \def\CO{{\cal O}}

    \def\CS{{\cal S}}

    \def\CV{{\cal V}}

  \def\({\left(}
    \def\){\right)}
    \def\<{\left\langle\,}
    \def\>{\, \right\rangle}
    \def\[{\left[}
    \def\]{\right]}
    \def\arccoth{\,{\rm arccoth}\,}
     \def\arccot{\,{\rm arccot}\,}

\def\pt{\partial}
\renewcommand{\Im}{\mathrm{Im}}

\title{Quantum folded string and integrability: from finite size effects to Konishi dimension}

\author{Nikolay Gromov \\Mathematics Department, King's College London,
      The Strand, London WC2R 2LS, UK\\ and\\ St.Petersburg INP, Gatchina, 188 300, St.Petersburg, Russia\\
    \email{nikolay.gromov AT kcl.ac.uk}}
\author{Didina Serban \\Institut de Physique Th\'eorique, DSM, CEA, URA2306 CNRS, Saclay, F-91191 Gif-sur-Yvette, France\\
    \email{didina.serban AT cea.fr}}
\author{Igor Shenderovich \\Institut de Physique Th\'eorique, DSM, CEA, URA2306 CNRS, Saclay, F-91191 Gif-sur-Yvette, France\\ and\\ St.Petersburg Department of
  Steklov Mathematical Institute RAS,\\ 27, Fontanka, 191023 St.Petersburg, Russia\\
    \email{shender.i AT gmail.com}}
\author{Dmytro Volin\\Department of Physics, The Pennsylvania State University,\\
University Park, PA 16802, USA\\and\\ Bogolyubov Institute for Theoretical Physics,\\ 14b Metrolohichna str, Kyiv 03680, Ukraine

\email{dvolin AT psu.edu}}
\maketitle

\abstract{ Using the algebraic curve approach we one-loop quantize the
  folded string solution for the type IIB superstring in $AdS^5\times
  S_5$.  We obtain an explicit result valid for arbitrary values of
  its Lorentz spin $S$ and R--charge $J$ in terms of integrals of elliptic functions. Then we consider the limit $S\sim J\sim 1$ and
  derive the leading three coefficients of strong coupling expansion
  of short operators. Notably, our result evaluated for the anomalous
  dimension of the Konishi state gives
  $2\lambda^{1/4}-4+2/\lambda^{1/4}$. This reproduces correctly the
 values predicted numerically in
  \href{http://xxx.lanl.gov/abs/0906.4240}{{\tt arXiv:0906.4240}}.
  Furthermore we compare our result using some new numerical data from
  the Y-system for another similar state. We also revisited some of the large
  $S$ computations using our methods. In particular, we derive
  finite--size corrections to the anomalous dimension of operators
  with small $J$ in this limit.  }

\preprint{KCL-MTH-11-03\\IPhT-T11/017}
\usepackage{varioref}

\begin{document}

\newpage

\section{Introduction}
In the last decade, enormous progress was made in computing the
spectrum of conformal dimensions of the ${\cal N}=4$ SYM theory in the
planar limit, which by the AdS/CFT correspondence
\cite{Maldacena:1997re,Witten:1998qj,Gubser:1998bc} is also the
spectrum of strings moving in $AdS_5 \times S^5$. This progress was
made possible by the discovery of integrability
\cite{Lipatov:1993yb,Faddeev:1994zg,Minahan:2002ve,Beisert:2003tq,Bena:2003wd,Kazakov:2004qf}
(also see review \cite{Beisert:2010jr} for further references). The
conjectured asymptotic Bethe ansatz equations
\cite{Beisert:2003yb,Beisert:2005fw,Beisert:2006ez} (also in the
review cited above) interpolating between weak and strong coupling
allowed to perform refined checks for operators with large
charges. During the last few years it became clear that the exact
solution for operators with finite charge is given by the Y-system
\cite{Gromov:2009tv,Bombardelli:2009ns,Gromov:2009bc,Arutyunov:2009ur,Cavaglia:2010nm}. At
strong coupling the Y-system was successfully tested in the
quasi-classical regime \cite{Gromov:2009tq,Gromov:2010vb}. The leading
weak coupling correction from the Y--system \cite{Gromov:2009tv}
agrees with direct perturbative computations
\cite{Fiamberti:2007rj,Velizhanin:2008jd}. The 5–loop corrections are
equivalent \cite{Arutyunov:2010gb,Balog:2010xa,Balog:2010vf} to the
L\"uscher corrections \cite{Bajnok:2008bm,Bajnok:2009vm} and also
consistent \cite{Kotikov:2007cy,Bajnok:2008qj,Lukowski:2009ce} with
the constraints from the BFKL equation \cite{Kotikov:2002ab}.

In \cite{Gromov:2009tv} the Y-system was combined with the vacuum TBA equations to produce an infinite set of integral equations for the         $sl(2)$ part of the
spectrum which were then solved numerically for the simplest
nontrivial Konishi \cite{Gromov:2009zb} operator. This method allowed
to find the anomalous dimensions of short operators in a completely
nonperturbative fashion starting from zero coupling and up to a
relatively large value of the `t Hooft coupling $\lambda$. For
extremely large $\lambda$'s the numerical solutions become very slow, the largest value of $\lambda$ reached up to now is
about $2000$ \cite{Frolov:2010wt}.  It is however possible to
extrapolate the numerical results and obtain the strong coupling
expansion of the anomalous dimensions.  The prediction obtained in
\cite{Gromov:2009zb} for the Konishi anomalous dimension $\gamma$ is
\beq\la{eq:eq} \gamma+4=2.0004\lambda^{1/4}+1.99/\lambda^{1/4}\;.
\eeq The leading coefficient agrees with the prediction of
\cite{Gubser:2002tv} giving $2$.  This was also confirmed in a recent paper \cite{Passerini:2010xc}. 
The sub-leading $2$ was in
disagreement with the calculation of \cite{Roiban:2009aa} which
appeared nearly the same time with \cite{Gromov:2009zb}. In \cite{Arutyunov:2009ax} it was
discussed that this disagreement could be attributed to singularities
which might change the integral equations at some large values of the coupling constant.  In the present paper we confirm analytically the
prediction (\ref{eq:eq}) and solve this long standing
discrepancy.

The strong coupling limit for the short operators looks very difficult
to address both within the integrability and the perturbative string
theory approaches. Although it is at least possible to define the set
of equations to be solved using the Y-system, it seems rather hard to
perform strong coupling expansion of these equations.  At finite
coupling the Y-functions appear to have a complicated analytical
structure \cite{Cavaglia:2010nm}, which can be already seen from its asymptotic solution \cite{Gromov:2009tv}.  It is likely that in general they have
infinitely many cuts and in the strong coupling
limit these cuts merge into each other.

On the string side, the semi-classical approach typically demands that
the conserved charges scale as the coupling constant.  For the
case of the spinning folded string the two charges, the Lorentz spin
$S$ and R-charge $J$, should scale in such a way that the ratios
$\CS=S/\sqrt{\l},\ \CJ=J/\sqrt{\l}$ remain fixed.  The expansion of
the energy is of the form \beq\label{enrgyexpnt} E\equiv
\gamma+S+J=\sqrt{\l}\,E_0(\CS,\CJ)+E_1(\CS,\CJ)+\frac
1{\sqrt{\l}}E_{2}(\CS,\CJ)+\ldots\ .  \eeq Then, in order to approach
the short operator regime we re-expand the
result \label{energyexpansion0} in the limit when $S,J\sim 1$ and thus
$\CS\sim \CJ\ll 1$.  As it was pointed out in \cite{Roiban:2009aa} the
expansion above reorganizes into a power series of the type \beq
E=\lambda^{1/4}a_0+\frac{1}{\lambda^{1/4}}a_2+\dots\;, \eeq where only
the classical energy $E_0$ contributes to the first coefficient $a_0$ and
both $E_1$ and $E_0$ contribute to the coefficient $a_2$. Thus with
some caution one may assume that the short strings with $S,J\sim 1$,
which are in principle deeply quantum states, still can be reached
using the quasi-classical methods. In this way, the complications with
the direct treatment of the Y-system can be escaped.  In this paper we
compute the first two coefficients in the expansion (\ref{enrgyexpnt})
using the algebraic curve quantization procedure for an arbitrary
$\CS$ and $\CJ$ (see \cite{SchaferNameki:2010jy} for more details).
What we found from our expressions for $E_0$ and $E_1$ is the
following expansion \beq\la{eq:SJ}
E=\lambda^{1/4}\sqrt{2S}+\frac{1}{\lambda^{1/4}}\frac{2J^2+S(3S-2)}{4\sqrt{2S}}+\dots\;.
\eeq Notice that this procedure is very straightforward and is free
from any ambiguity. What is important is that we do not get any
logarithmic terms which would signal order-of-limits problems.  From
the ABA we know that the Konishi state in the $sl(2)$ sector is given
by $S=J=2$.  Substituting these values of the parameters we
indeed obtain a result consistent with the prediction of \cite{Gromov:2009zb} \eq{eq:eq}.  In
order to rule out an accidental coincidence of our result with that of
\cite{Gromov:2009zb} in the case of the Konishi operator, we also made a comparison of our result
\eq{eq:SJ} with the numerical data obtained by \cite{Gromov:2011xx}
for another similar state, with $S=2 \ J=3$.

The limit of short strings would not be the only case when the
analytical continuation of the algebraic curve results gives reliable
results in regions where the L\"uscher corrections are large and the
Y-system is difficult to handle analytically. Another example, which
is treated in the second part of this paper, is that of the long
strings with large Lorentz spin $S$ and small twist $J=\ell \,4g \log
S$.  It is well understood that this case can be obtained from the
generic two-cut solution in the $sl(2)$ sector
\cite{Frolov:2006qe,Casteill:2007ct,Belitsky:2007kf,Gromov:2008en}
\be\label{logSexpansion}
E(\ell,\CS)=\(\sqrt{\l}f_0(\ell)+f_1(\ell)+\frac
1{\sqrt{\l}}f_2(\ell)\)\log \CS+\ldots\,.  \ee The leading logarithmic
scaling is a generic feature in gauge theories
\cite{Gross:1974cs,Korchemsky:1988si,Korchemsky:1992xv,Belitsky:2006en}
and the coefficient of $\log \CS$ is the so-called generalized scaling
function. The functions $f_0(\ell)$, $f_1(\ell)$, $f_2(\ell)$ were
derived explicitly in
\cite{Frolov:2006qe},\cite{Frolov:2006qe,Casteill:2007ct,Belitsky:2007kf},\cite{Gromov:2008en,Volin:2008kd,Giombi:2010fa}
respectively. It happens that all the three coefficients have a
well-defined limit at $\ell=0$ and that in this limit they reproduce
correctly the strong coupling expansion of the cusp anomalous
dimension. In particular the $\ell=0$ result obtained in this order of
limits coincides with the solution obtained
\cite{Kotikov:2006ts,Benna:2006nd,Alday:2007qf,
  Kostov:2007kx,Beccaria:2007tk,Basso:2007wd,Kostov:2008ax} via the
BES equation \cite{Beisert:2006ez,Eden:2006rx}, which supposes $J=2$ and $S$ large
and then $g\to\infty$. A recent review of this subject appeared in
\cite{Freyhult:2010kc}.

In this paper we also revisit the computation of the classical and
one-loop energy for the long string both from the point of view of the
algebraic curve and the Y-system, having in view the finite size
corrections. We obtain results for all orders in $1/\log S$ and we
neglect terms of the order $\log S/S$ and higher. At $\ell=0$ the
result is particularly simple\footnote{Here we use the alternative
  notation $g=\sqrt{\lambda}/4\pi$.}
\begin{equation}
\label{eq:lslo}
E_{\ell=0}=S+J+ 4g\(\log \frac{2S}{g}-1\)-\frac{3\log 2}{\pi}\log \frac{2S}{g}+\frac{6\log 2}{\pi}+1-\frac{5\pi}{12\log (2S/g)}+\CO(1/g)
\end{equation}
The sub-leading part in $\log S$ is the so-called virtual scaling function computed in \cite{Freyhult:2009my,Fioravanti:2009xt} while the $1/\log S$ part agrees with the results in \cite{SchaferNameki:2005is,Beccaria:2010ry,Giombi:2010zi}.
In  \cite{Giombi:2010zi} the last term in (\ref{eq:lslo}) was given the simple interpretation of
contribution of massless excitations propagating on a string of length $L=2\log S$, with total result:
\be
        \delta E_1=-\frac {\pi}{12\log S}\times(\rm number\ of\ massless\ modes).
\ee
The massive modes lead to correction of the type $e^{-mL}$, where  $m\sim\ell$ and $L=2\log S$; these contributions have to be summed up properly in order to reproduce the massless limit.

In our computation, the four massive mode contributions come via the wrapping corrections. From the Y-system point of view, this part is constituted by two contributions (virtual particle contribution and back-reaction of the roots) which become separately divergent when $\ell\to 0$. The algebraic curve computation does not see any divergence, and this may be compared to the particularly smooth behavior of the algebraic curve prediction for the short strings.

Finally, the contribution of the massless mode comes via the
asymptotic Bethe ansatz. This might seem surprising, since at finite
coupling there are no $1/\log S$ corrections for the twist-two
operator $J=2$, just the $(\log S)^0$ term
\cite{Basso:2006nk,Fioravanti:2009ei}. This is obviously due to the
different order of limits which are considered and might be explained
by the fact that the bosonic modes of the $O(6)$ sigma model \cite{Alday:2007mf,Basso:2008tx,Fioravanti:2008rv} acquire a
dynamically generated mass at finite coupling.\footnote{We have been informed by B. Basso that this aspect will be investigated in \cite{Basso:2011xx}.}

In conclusion, the algebraic curve method is a very reliable and
efficient tool to obtain the one-loop results for a various range of
string solution and seems to be free of some of the difficulties
inherent to the direct treatment of the Y-system at strong coupling as
well as from the ambiguities of the direct worldsheet quantization.
The algebraic curve method may serve as a starting point to understand
the behavior of the Y-system at higher loop order.

The plan of the paper is as follows. In section 2 we present the
essential data for the algebraic curve which is necessary to derive
the one-loop energy for the folded string at arbitrary $\CS$ and
$\CJ$. In subsection 2.4 we specialize to the operators with $S$ and
$J$ finite, including the Konishi operator, and we
compare the result with the available numerical predictions. In
section 3. we compute the spectrum for the long string to one loop,
first from the algebraic curve and then from the Y-system.

\section{Folded string quasi-classical quantization from Algebraic Curve}
\label{sec:folded}
The algebraic curve method is one of the most advanced ways of
computing the semi-classical corrections in the AdS/CFT
correspondence. The method is heavily based on integrability and is
naturally free from the usual perturbation theory ambiguities.  Some
of these ambiguities are listed in Appendix E of \cite{Gromov:2007aq}.
Some of the question raised there were recently explored in more
detail in \cite{Mikhaylov:2010ib}.

The method of the quasi-classical quantization was recently very
pedagogically described in the recent AdS/CFT integrability review
chapter \cite{SchaferNameki:2010jy} were more references can be
found. We will adopt the same notations here.

\subsection{Classical solution}
The only input needed to proceed with the quantization is a set of
quasi-momenta $\hat p_i,\tilde p_i,\;i=1,2,3,4$ which constitute the
algebraic curve.  The folded string solution corresponds the curve
with two symmetric cuts with real branch-points $\pm b, \pm a$ such
that $1<a<b$. The explicit form of the quasi-momenta depends on the
twist $J$ and the Lorentz spin $S$ can be constructed using the
methods of \cite{Beisert:2003ea,Kazakov:2004qf,Kazakov:2004nh}. What one finds are
the following expressions \beqa
  \label{eq:p_a}\nn
  p_{\hat 2} &=& \pi n - \frac{J}{2g} \left( \frac{a}{a^2-1} -
    \frac{x}{x^2-1} \right) \sqrt{\frac{(a^2-1)
      (b^2-x^2)}{(b^2-1)(a^2-x^2)}} \\ \nn &+& \frac{2 a b S F_1(x)}{g(b-a)(ab+1)} +
  \frac{J (a-b) F_2(x)}{2g\sqrt{(a^2-1)(b^2-1)}},\\
  \label{eq:p_s}
  p_{\tilde 2} &=& \frac{Jx}{2g(x^2-1)}.
\eeqa
The integer $n$ (the mode number) is related to the number of spikes and
$g=\frac{\sqrt{\lambda}}{4\pi}$.
All the other quasi-momenta can be
found from the standard symmetry relations for the $sl(2)$ sector
\beqa
  \label{eq:quasimomenta_symmetry_A}
  p_{\hat{2}} (x) &=& -p_{\hat{3}}(x) = -p_{\hat{1}}(1/x) = p_{\hat{4}}
  (1/x)\;,\\
  \label{eq:quasimomenta_symmetry_S}
  p_{\tilde{2}} (x) &=& -p_{\tilde{3}} (x) = p_{\tilde{1}} (x) =
  -p_{\tilde{4}}(x)\;.
\eeqa
The functions $F_1(x)$ and $F_2(x)$ can be expressed
in terms of the elliptic integrals:
\begin{eqnarray}
 Ê\label{eq:not}
 ÊF_1(x) &=& i F \left( i \sinh^{-1}
 Ê Ê\sqrt{\frac{(b-a)(a-x)}{(b+a)(a+x)}} | \frac{(a+b)^2}{(a-b)^2} \nonumber
 Ê\right)\;, \\
\nn
 ÊF_2(x) &=& i E \left( i \sinh^{-1}
 Ê Ê\sqrt{\frac{(b-a)(a-x)}{(b+a)(a+x)}} | \frac{(a+b)^2}{(a-b)^2}
 Ê\right)\;.
 Ê\end{eqnarray}
The branch points $a,b$ are fixed once $S$ and $J$ are specified by the following equations
\beqa
 Ê\label{eq:global_s}
 ÊS &=& 2ng \frac{ab+1}{ab} \left(b E \left( 1- \frac{a^2}{b^2} \right)
 Ê Ê-a K \left( 1- \frac{a^2}{b^2} \right) \right),\\
 Ê\label{eq:global_l}
 ÊJ &=& \frac{4ng}{b} K \left( 1- \frac{a^2}{b^2} \right) \sqrt{(a^2-1) (b^2-1)}\;.
\eeqa
Then the classical energy can be computed from
\begin{equation}
 Ê\label{eq:global_delta}
 Ê\Delta = 2ng \frac{ab-1}{ab} \left( b E \left( 1- \frac{a^2}{b^2}
 Ê Ê\right) + a K \left( 1- \frac{a^2}{b^2} \right) Ê\right).
\end{equation}
In this way we get the classical energy of the generalized-folded solution
as a function of $S$ and $J$.

It is important to notice that there is no question about
identification of the operators corresponding to this classical
solution in the framework of the algebraic curve. The algebraic curve
was mapped to the Asymptotic Bethe Ansatz in
\cite{Kazakov:2004qf,Beisert:2005bm}.  The super Yang-Mills operators
at one loop are in one to one correspondence with the Bethe ansatz
solutions up to a global bosonic symmetry action. Thus we can identify
the class of operators we consider here. It is a subset of ${\rm
  tr}(D^S Z^J)+{\rm perm.}$ Basically, since we use the algebraic
curve formalism, we know automatically the whole set of conserved
charges which provides an exhaustive information about the state.  In
particular this set includes the Konishi operator when $J=2,S=2$ and
the mode number $n$ chosen to be $n=1$ as it follows from the one-loop
spectrum at weak coupling.

\subsection{Off-shell fluctuations}
An important feature of the algebraic curve quantization is that one
can work with the off-shell fluctuation as it is described in detail
in \cite{SchaferNameki:2010jy}. The off-shell fluctuation energies as
functions of the spectral parameter $x$ are much simpler than the
usual fluctuation energies, usually obtained in the world-sheet
quantization procedure, which are functions of mode numbers.  The
former should coincide with the later when evaluated at the special
points of the curve given by \beq\la{eq:flpz}
p_i(x_k^{ij})-p_j(x_k^{ij})=2\pi k\;.  \eeq In general one has to
compute $8+8$ different off-shell energies corresponding to the number
of the physical world-sheet degrees of freedom.  However as it was
shown in \cite{Gromov:2008ec} in the rank one sectors one can express
all of them in terms of just two, $\Omega^{\hat 2\hat 3}$ and
$\Omega^{\tilde 2\tilde 3}$:
\begin{eqnarray}\nn
  \label{eq:frequencies}
  \Omega^{\hat{1}\hat{4}} (x) &=& -\Omega^{\hat{2}\hat{3}}(1/x) -2\;,\;\;\\
  \Omega^{\hat{1}\hat{3}} (x) &=& \Omega^{\hat{2} \hat{4}} (x) = \frac12
  \Omega^{\hat{1}\hat{4}}(x) - \frac12 \Omega^{\hat{1}\hat{4}}(1/x) -1\;,\nn\\
  \Omega^{\hat{1}
    \tilde{3}}(x)
&=&
  \Omega^{\hat{1}\tilde{4}}(x)
    =  \Omega^{\hat{4}\tilde{1}}(x)
    =  \Omega^{\hat{4}\tilde{2}}(x)
     =
  \frac12\, \Omega^{\tilde{2}\tilde{3}} (x) + \frac12\,\Omega^{\hat{2}\hat{3}}(x),\\
  \Omega^{\hat{2}\tilde{3}}(x)
&=&
  \Omega^{\hat{2}\tilde{4}}(x)
=
  \Omega^{\hat{3}\tilde{1}}(x)
=
  \Omega^{\hat{3}\tilde{2}}(x)
= \frac12\, \Omega^{\tilde{2}\tilde{3}}(x) - \frac12\, \Omega^{\hat{2}\hat{3}}(1/x)-1,\nn\\
  \Omega^{\tilde{2}\tilde{3}}(x)
&=&
  \Omega^{\tilde{2}\tilde{4}}(x)
=
  \Omega^{\tilde{3}\tilde{1}}(x)
=
  \Omega^{\tilde{3}\tilde{2}}(x)
= \Omega^{\tilde{2}\tilde{3}}(x)\;.\nn
\end{eqnarray}
Since we are considering the $sl(2)$ sector, the fluctuation energies in $S^5$
should be trivial and can be written down immediately:
\beq
\label{oma}
\Omega^{\tilde 2\tilde 3}(x)=+\frac{2}{a
  b-1}\frac{\sqrt{a^2-1}\sqrt{b^2-1}}{x^2-1}\;.  \eeq Calculation of
$\Omega^{\hat 2\hat 3}(x)$ is a little bit more involved. However the
steps one should follow are exactly the same as in
\cite{Gromov:2008ec} and we simply give the result here \beqa
\label{oms}
\Omega^{\hat 2\hat 3}(x)=+\frac{2}{a b-1}\(1-\frac{y(x)}{x^2-1}\)\;,
\eeqa
where $y(x)=\sqrt{x-a} \sqrt{a+x} \sqrt{x-b} \sqrt{b+x}$.

For the analytical properties of these fluctuation energies see
\cite{Gromov:2008ec}.

\subsection{One-loop shift}
\label{sec:1-loop}

In the previous sections we prepared all necessary ingredients needed
for the one-loop corrections to the classical energy. As we mentioned
in the previous section the usual excitation energies, typically used
in the worldsheet calculations, can be obtained from the off-shell
fluctuation energies $\Omega^{ij}(x)$ by setting $x$ to the value
given be the equation \eq{eq:flpz} and then sum over all polarizations
$(ij)$ and all mode numbers $k$. Doing this explicitly is almost
impossible for the given quasi-momenta.  The standard way to overcome
this difficulty is to rewrite the sum as an integral (see, for
example, \cite{Gromov:2008ec}). For precise contour description see,
for example, \cite{Gromov:2007cd}.

\begin{equation}
  \label{eq:1-loop}
  \mathcal{E} = \frac12 \sum_{ij} (-1)^{F_{ij}} \oint \frac{dx}{2\pi
    i} \left( \Omega^{ij}(x)\, \pt_x \log \sin \frac{p_i-p_j}{2} \right).
\end{equation}
Here $F_{ij}$ is the fermionic number: $F_{ij} = 0$ for bosonic
polarizations and $F_{ij}=1$ for fermionic. The term
$\partial_x\log \sin \frac{p_{i}-p_j}{2}$ has the poles at the solutions of
\eq{eq:flpz}.
 The contour of
integration encircles all the possible fluctuations
$x^{ij}_k$.
This result is already explicit enough, however it is instructive to deform
the contour into the unit circle (for each $(ij)$).
During this contour deformation we can get two types of terms:

\begin{itemize}
\item Contribution from the integration on the  unit circle  for each
  polarization $(ij)$;
\item Additional contribution from the cuts of the classical
  solution. Only the term with $(ij)=(\hat 2\hat 3)$
  gain such contribution.
\end{itemize}
It is also convenient to use the variable $z$ instead of $x$:
\begin{equation}
  \label{eq:zhukovsky}
  x = z + \sqrt{z^2-1},
\end{equation}
which maps the unit circle $|x|=1$ onto the interval $z\in[-1,1]$.
Also we can split the logarithm in
two parts:
\begin{equation}
  \label{eq:log_split}
  \log \sin \frac{p_i-p_j}{2} = \frac{i(p_i-p_j)}{2} +
  \log \left( 1 - e^{-i(p_i-p_j)} \right).
\end{equation}
which holds up to some irrelevant constant. In this way we split the finite
size effects from the asymptotic contribution. Indeed, for $z\in[-1,1]$
 $e^{-i(p_i-p_j)}$ is exponentially suppressed for large $J$.
Substituting this into \eqref{eq:1-loop}, we get two terms, $\delta
E_1$ and $\delta E_2$:

\begin{equation}
  \label{eq:delta_E_1}
  \delta E_1 = \sum_{ij} (-1)^{F_{ij}} \int\limits_{-1}^{1} \frac{dz} {2\pi
    i} \left( \Omega^{ij}(z) \, \pt_z \frac{i (p_i -p_j)}{2} \right),
\end{equation}

\begin{equation}
  \label{eq:delta_E_2}
  \delta E_2 = \sum_{ij} (-1)^{F_{ij}} \int\limits_{-1}^{1} \frac{dz} {2\pi
    i} \left( \Omega^{ij}(z) \, \pt_z \log (1- e^{-i(p_i -p_j)}) \right).
\end{equation}
One should take in account the contribution which we get by
deforming the contour, which encircles the cuts $[\,-b,-a\,]$ and
$[\,a,b\,]$. This contribution can be written as
\begin{equation}
\label{eq:delta_E_3}
  \delta E_3 = -\frac{4}{ab-1}\int_a^b \frac{dx} {2\pi i}
  \frac{y(x)}{x^2-1} \partial_x \log \sin p_{\hat 2}.
\end{equation}
where we use \eqref{eq:quasimomenta_symmetry_A},
\eqref{eq:quasimomenta_symmetry_S}.
The one-loop shift is then given by
\beq
E^{\rm1-loop}=\delta E_1 + \delta E_2 + \delta E_3\;.
\eeq
In Appendix \ref{AppA} we further expanding these integral using the
relations between frequencies with different polarizations $(ij)$.

\subsection{\label{sec:shortoperatorlimit}Short operator limit}
In this section we will exploit the explicit exact result for one-loop
applicable for arbitrary $J,S\sim g$ which was derived in the previous
section. These formulae involve a single integration and they can be
evaluated numerically for various values of parameters.

The analytical evaluation of these integrals in general is not
straightforward.  In some limits, however, the integrands could
simplify considerably so that the integration can be performed
analytically. In this section we will consider one of such limits,
namely we fix the ratio $r=J/S$ and then expand the result for small
$S/g$.  We will then motivate the relevance of this limit for the
Konishi operator as well as for the similar type of operators with
very few fields.

This limit is not completely trivial,
the reason being that the algebraic curve becomes singular in this case:
both positive branch points $a$ and $b$ approach the pole at $x=1$
as it can be easily seen from \eq{eq:global_s} and \eq{eq:global_l}.
We denote
\beq
s\equiv\frac{\sqrt{2S/n}}{\lambda^{1/4}}\;\;,\;\;r\equiv \frac{J}{S}\;,
\eeq
where ${\sqrt\lambda}=4\pi g$. In these notations
the expansion of \eq{eq:global_s} and \eq{eq:global_l} gives
\beqa
a&=&1+\frac{r^2 }{8}s  ^3+\frac{r^2-r^4}{128} s  ^5+\frac{r^4}{128}s ^6-\frac{9 r^2+22 r^4-4 r^6}{4096}s  ^7+\dots\;,\\
b&=&1+2 s  +2 s  ^2+\frac{7+r^2}{8} s  ^3-\frac{1-r^2}{4}s  ^4-
\frac{85-34 r^2+2 r^4}{256}s  ^5+\dots \;.
\eeqa
The classical energy gives
\beq\la{re:clas}
\frac{\Delta}{ n \sqrt\lambda}=s +\frac{3+2 r^2}{16}s^3-\frac{21-20 r^2+4 r^4}{512}
 Ês ^5+O\left(s ^7\right)+\dots \;.
\eeq
We consider in some detail only the evaluation of $\delta E_1$ and then give the result for the others
integrals. In what follows we restrict ourselves to the case $n=1$. In general, the expression for $\delta E_1$ can be written conveniently as it is shown in App.\ref{AppA}:
\beq
\delta E_1=-\frac{2}{\pi}\int_0^{1} {\rm Im}\(\Omega^{\hat 2\hat 3}(z)-\Omega^{\tilde 2\tilde 3}(z)\) {\rm Im}\left(
 Ê p_{\hat 2}'(z)-p_{\tilde 2}'(z)\right)dz\;.
\eeq
First, assuming $z-1\sim 1$ we expand the integrand to get
\beq
-\int_0^1\frac{2z^2 s}{(z^2-1)^2}dz+\dots\;.
\eeq
Apparently the integral is divergent close to $z=1$.
This divergence should be canceled when the integrand is treated more
accurately for small $z-1$. There are two important scales when $z$
approaches $1$: when one zooms close to the branch point $b$ which
scales as $s^2$ then $z=1-s^2\zeta$ and when one further zooms so that
we can distinguish the smallest branch point $a$ from $1$
i.e. $z=1-s^6\xi$. For each of these scales the integral is divergent,
however, when all the three regions are combined together the
divergences must cancel. What we get for $\delta E_1$ is \beq
\delta E_{1}\simeq Ê -s\log \frac{rs^2}{2} -\frac{s }{2}\,.\\
\eeq The contributions $\delta E_2$ and $\delta E_3$ can be computed
similarly.  The results for these contributions are \beqa \delta
E_2&\simeq&s Ê\log s+
c_1 s\,,\\
\delta E_3&\simeq&s\log \frac{r s}{2}+\frac{s}{4}-c_1s\,, \eeqa where
the constant $c_1\simeq 0.0203628454$. Notice that there are various
$\log$ divergences which all cancel when the terms are combined
together and the final result is very simple\footnote{ In order to
  get the ABA result with Hernandez-Lopez phase one should drop the
  $\delta E_2$ contribution. In this case one would get $-s\log
  s\simeq \frac{\sqrt{2S}}{4\lambda^{1/4}}\log\lambda$ divergence.
  Exactly this divergence was indeed observed in \cite{Rej:2009dk} for
  the Konishi anomalous dimension ($S=2,n=1$) computed in the ABA
  framework.  } \beq\la{re:1loop} \Delta^{\rm 1-loop}=\delta
E_1+\delta E_2+\delta E_3\simeq-\frac{s}{4}
=-\frac{\sqrt{2S}}{4\lambda^{1/4}} \;.  \eeq In fact this result was
previously obtained by us numerically with only two digits
precision. This unpublished result was already used in
\cite{Roiban:2009aa} for the Konishi operator.  The main difference
with \cite{Roiban:2009aa} is that we do not assume that $J=0$, instead
from the point of view of algebraic curve and its relation to the ABA
it is rather obvious that one should take $J=2$ for Konishi operator
instead.  Here we follow the approach of \cite{Roiban:2009aa} with
this small modification, which, however, changes the result
considerably\footnote{Almost simultaneously with us this point was
  also realized by A.Tseytlin according to our private
  communication.}.  Now we simply combine the classical energy
\eq{re:clas} and the one-loop result \eq{re:1loop} to get
\beq\la{eq:loopres} \Delta^{\rm classical}+\Delta^{\rm
  1-loop}=\lambda^{1/4}\sqrt{2S}
+\frac{1}{\lambda^{1/4}}\frac{2J^2+S(3S-2)}{4\sqrt{2S}}\;.  \eeq The
contribution to the first term comes solely from the classical energy,
whereas both classical energy and the one-loop energy contribute to
the second term.  It is very tempting to assume that this pattern
will continue further and in order to find the contribution to the
next term one should also compute a two-loop
correction\footnote{Strictly speaking this result holds assuming no
  non-perturbative terms contribute and that at each loop level the
  contribution can be represented as a regular series in $s$ vanishing
  at $s=0$.}.

\FIGURE{\la{fig321}
  \centering
  \includegraphics{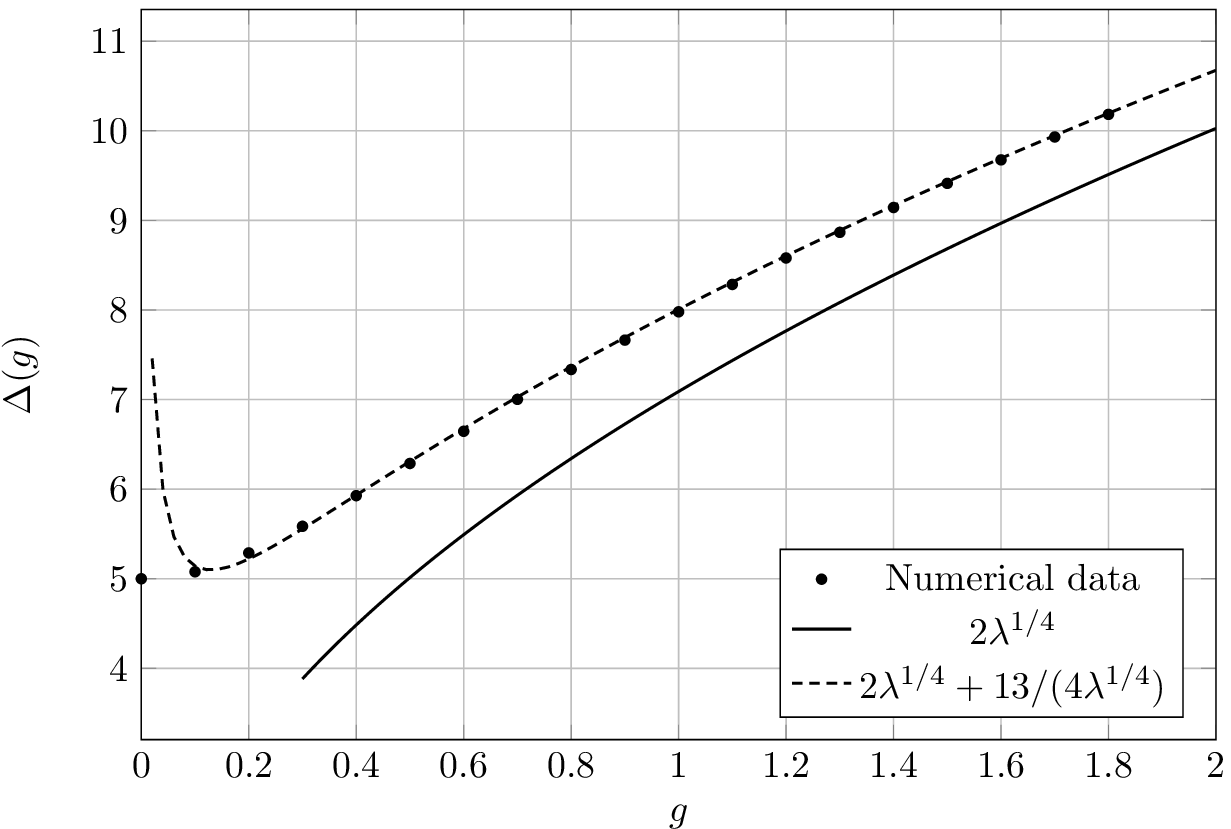}
  \caption{Numerical results from the Y-system for $J=3,\ S=2,\ n=1$ compared with our analytical strong coupling expansion (\ref{eq:loopres}). Here as everywhere in the paper $\lambda=16\pi^2 g^2$.}
}
\TABLE{\label{tb:data}
\begin{tabular}{|l|l||l|l|}
 \hline

 $g$ & $\Delta$ & $g$ & $\Delta$\\   \hline
 0 & 5 & 0.9 & 7.6632\\
 0.1 & 5.0777 & 1.0 & 7.9794\\
 0.2 & 5.2883 & 1.1 & 8.2848\\
 0.3 & 5.5854 & 1.2 & 8.5801\\
 0.4 & 5.9275 & 1.3 & 8.8661\\
 0.5 & 6.2868 & 1.4 & 9.1436\\
 0.6 & 6.6456 & 1.5 & 9.4129\\
 0.7 & 7.0023 & 1.6 & 9.6752\\
 0.8 & 7.3354 & 1.7 & 9.9308\\
 \hline
\end{tabular}
\caption{Konishi-like operator with $J=3$.
The full dimension $\Delta=\gamma^{\rm anom}+S+J$ for various values of $g$.
Numerical data by \cite{Gromov:2011xx}
obtained with a new exact truncation method \cite{Kazakov:2009stock}.
The numerical absolute error is about $\pm 3\times 10^{-4}$.
}
}
The equation \eq{eq:loopres} produces an infinite set of predictions which can be verified using the
Y-system numerical approach proposed by \cite{Gromov:2009zb}.

For the Konishi state the prediction
is already available \cite{Gromov:2009zb}. To compare one should substitute $S=J=2$
 as we discussed above. Eq (\ref{eq:loopres}) produces
\beq
\left.\Delta^{\rm classical}+\Delta^{\rm 1-loop}\right|_{S=2,J=2}=2\lambda^{1/4} +\frac{2}{\lambda^{1/4}}\;
\eeq
in the perfect agreement with the Y-system prediction of \cite{Gromov:2009zb}\footnote{It is also compatible with the
recent numerics with slightly higher precision \cite{Frolov:2010wt}.}.

A natural question one can ask is whether this prediction is going  to be correct for short operators  other than Konishi.
To address this question we consider an operator similar to Konishi, with $J=3$, $S=2$,
which we denote as $(3,2,1)$.
From (\ref{eq:loopres}) we see that our prediction produces

\beq \Delta^{(3,2,1)}=2\lambda^{1/4} +\frac{13}{4\lambda^{1/4}}\;.
\eeq We compared this result with the preliminary numerical data
shared with us by the authors of \cite{Gromov:2011xx} (see
Tab.\ref{tb:data}).  As one can see from Fig.\ref{fig321}, our
analytical results perfectly match these numerical points\footnote{Let
  us also mention that our result \eq{eq:loopres} for $J=4,S=2,n=1$
  gives $2\lambda^{1/4}+5/\lambda^{1/4}$ which seems to be consistent
  with some of the preliminary numerical results reported in
  \cite{Frolov:2010talk} even though no solid statement was made
  there.  }.

\section{The $1/ \log S$ corrections for the long folded string}

In this section we derive the finite size corrections for the regime
when $S$ is large and $J=4g \ell \log S$ with $\ell$ finite.  The
corrections are obtained by computing the three integrals
(\ref{eq:delta_E_1})-(\ref{eq:delta_E_3}) (subsection
\ref{sec:subseqtionfromcurve}) and in an alternative way by using the
Y-system at one loop derived in \cite{Gromov:2009tq} (subsection
\ref{sec:subseqtionfromY}). We obtain the corrections at arbitrary
order in $1/\log S$, and we neglect all the inverse powers of $S$, as
well as the $\log S /S$ terms.  As a byproduct, we are re-deriving the
known results for the generalized scaling function up to one loop
\cite{Casteill:2007ct, Belitsky:2007kf}, as well as the virtual
scaling function \cite{Freyhult:2009my} to the same order. The
computations are done for arbitrary $\ell$, but of course the formulas
greatly simplify for the GKP \cite{Gubser:2002tv} limit $\ell=0$.  In
this limit, the energy is given by
\begin{equation}
  E_{\ell=0}=S+J+ 4g\(\log \frac{2S}{g}-1\)-\frac{3\log 2}{\pi}\log \frac{2S}{g}+\frac{6\log 2}{\pi}+1-\frac{5\pi}{12\log (2S/g)}+\CO(1/g)\;.
\end{equation}
The $(\log S)^0$ part is in agreement\footnote{The term ${6\log
    2}/{\pi}+1$ was previously known by one of us, see the note added
  on pg. 24 in \cite{Beccaria:2008tg}.  Please refer to the current
  paper for this coefficient.  } with \cite{Freyhult:2009my,Fioravanti:2009xt,Fioravanti:2010qh,Beccaria:2010ry}, while the $1/\log S$ part agrees with the
results in
\cite{SchaferNameki:2005is,Beccaria:2010ry,Giombi:2010zi}. The
$-{5\pi}/{12\log S}$ term can be interpreted as coming from the finite
size corrections associated to $5$ massless bosonic fields
\cite{Giombi:2010zi,Basso:2011xx}.  At $\ell\neq 0$, four of these
bosonic modes are massive, and their contribution is captured by the
wrapping corrections. The fifth mode is massless, and it contributes
via the anomaly term in the asymptotic Bethe ansatz
equations. Although at weak coupling the asymptotic Bethe ansatz
yields no $1/\log S$ corrections for the twist-two operator $J=2$,
\cite{Fioravanti:2009ei}, at strong coupling the situation is
different. This can be attributed to the different order in which the
limits $S\to \infty$ and $g\to\infty$ are taken.

\subsection{\label{sec:subseqtionfromcurve}The one-loop corrections for the long string from algebraic curve}

In the limit of the long string $S\to \infty$, the endpoints $\pm b$
of the cuts of the curve go to infinity and the solution becomes
effectively one-cut. The expression for the charges
(\ref{eq:global_s})-(\ref{eq:global_delta}) simplify and, up to
negative powers in $S$, we have
\begin{equation}
 \frac{S}{2g}=  b ,
\qquad
 \frac{J}{4g} =  \sqrt{a^2-1 } \log\frac{2S}{ag}\;,
\qquad
 \frac{ \Delta}{2g} =   \frac{S}{2g}
  + a \log\frac{2S}{ag} .
\end{equation}
In particular, we notice that the parameter $\ell\equiv J/4g\log S$ is related to the position of the endpoint $a$ of the cut by
\begin{equation}
\label{ella}
\ell =\sqrt{a^2-1}\(1-\frac{\log(ag/2)}{\log S}\)\;.
\end{equation}
There will be trivial $1/\log S$ terms coming from this relation between $\ell$ and $a$. After some manipulation, the elliptic functions reduce in the large spin limit to simpler functions
and the quasi-momenta (\ref{eq:p_a}) become
\begin{eqnarray}
 \label{eq:p_asim}
&\ &  p_{\hat 2}(x) = \frac{J}{2g\sqrt{a^2-1}} \frac{x \sqrt{a^2-x^2}
    }{x^2-1} -4\arctan\sqrt{\frac{a-x}{a+x}},\\ \nn
  &\ &   p_{\tilde 2}(x) = \frac{J}{2g} \frac{x
    }{x^2-1}\;.
    \end{eqnarray}
    The off-shell frequencies (\ref{oma}) and (\ref{oms}) are given in this limit by
    \begin{eqnarray}
 \Omega^{\tilde 2\tilde 3}=\frac{2}{a} \frac{\sqrt{a^2-1}}{x^2-1}\;, \quad    \Omega^{\hat 2\hat 3}=\frac{2}{a} \frac{\sqrt{a^2-x^2}}{x^2-1}\,.
    \end{eqnarray}
With these data in hand we are able to proceed to the computation of the three integrals giving the complete one-loop contribution to the energy.
The easiest part to compute is $\delta E_2$, which can be reduced to \begin{eqnarray}
   \label{massmodesAC}
 \delta E_{2}=\frac{4\sqrt{a^2-1}}{a\pi}\int_{1}^\infty dt \; \frac{\log (1-e^{-J t/2g})}{\sqrt{1-1/t^2}}\equiv \frac{\sqrt{a^2-1}}{a}\; \CI(2\ell \log S)\;.
 \end{eqnarray}
The following two representations of  $\CI(\alpha)$ are particularly useful:
\begin{eqnarray}
\label{expnI}
&\ & \CI(\alpha)=-\sum_{n=1}^\infty\frac{4}{n\pi} K_1(n\alpha)\\
 &=&-\frac{2\pi}{3\a}+2+\frac{\a}{2\pi}\left(2\gamma_E-1+2\log\left(\frac{\a}{4\pi}\right)\right)+\frac 1\pi\sum_{k=1}^\infty \frac{(-1)^k\zeta(2k+1)\Gamma(2k+1)}{\Gamma(k+1)\Gamma(k+2)(4\pi)^{2k}}\a^{2k+1}\,. \nn
 \end{eqnarray}
From the first representation\footnote{A similar representation was obtained by B. Basso \cite{Basso:2011xx}} we deduce that at large $\alpha$, $\CI(\alpha)\sim e^{-\alpha}/\sqrt{\alpha}$, so for finite $\ell$ the associated finite-size corrections vanish exponentially. The second representation in (\ref{expnI}) is useful in the
 small $\ell$ regime, where it gives\footnote{Note that (\ref{massmodesel}) appears only for $\ell\log S\ll 1$ which practically corresponds to  $\CJ\to 0$ limit prior to the large $\CS$ limit.}
\begin{eqnarray}
   \label{massmodesel}
 \delta E_{2}\simeq \frac{\ell \CI(2\ell\log S)}{\sqrt{1+\ell^2}}=-\frac{4\pi}{12\log S}+\CO(\ell)\;.
 \end{eqnarray}

 In the $O(6)$ language,
 \cite{Alday:2007mf},\cite{Basso:2008tx}, this is the correction
 coming from four of five bosonic modes which are massive at finite
 $\CJ$ (hence exponential suppression at large $\a$) but become
 perturbatively massless in the limit $\CJ\to 0$.

 The other two contributions to the one-loop energy are
 \begin{eqnarray}
\label{mainc}
\delta E_1&=&-\frac{4}{a}\int_{U_+}\frac{dy}{2\pi}{\,\rm Im\,} \frac{\sqrt{a^2-y^2}-\sqrt{a^2-1}}{y^2-1}\partial_y{\,\rm Im\,} G_0(y)\;,\nn \\
\delta E_3&=&-\frac{4}{a}\int_a^\infty \frac{dy}{2\pi }\frac{\sqrt{y^2-a^2}}{y^2-1}p'\coth p\;,
\end{eqnarray}
where we have denoted $p\equiv i p_{\hat 2}(y+i0)$, $G_0(y)\equiv  p_{\hat 2}(y)-p_{\tilde 2}(y)$ and the contour $U_+$ is the upper half of the unit circle running clockwise.
The last  term can be split naturally into two parts
\begin{eqnarray}
  \delta E_3=\delta E_{3,{\rm an}}+ \delta E_{3,{\rm m}}\;.
\end{eqnarray}
The anomaly-like term $\delta E_{3,{\rm an}}$  contains the finite-size corrections associated to the fifth bosonic mode, which remains massless for arbitrary $\ell$
\begin{eqnarray}
 \label{E1an}
\delta E_{3,{\rm an}}\equiv -\frac{4}{a}\int_a^\infty \frac{dy}{2\pi }\frac{\sqrt{y^2-a^2}}{y^2-1}p'(\coth p-1)=-\frac{\pi}{12(1+\ell^2)  \log S}+\CO(1/\log^2 S)\,.
\end{eqnarray}
This is exactly the contribution of the only at $\CJ\neq 0$ massless
mode identified by Giombi, Ricci, Roiban and Tseytlin
\cite{Giombi:2010zi}. A more refined evaluation of the anomaly part, up to
$\log S/S$ terms, can be straightforwardly done using
\begin{eqnarray}
 \label{E1ancomp}
\delta E_{3,{\rm an}}= \sum_{n=1}^\infty f^{(n)}(0)\frac{\zeta(n+1)}{2^n}+\CO(\log S/S) \quad {\rm with}\quad    f(p)=-\frac{2}{a\pi }\frac{\sqrt{y^2(p)-a^2}}{y^2(p)-1}.
\end{eqnarray}
The remaining two contributions $\delta E_{1}$ and $\delta E_{3,{\rm m}}$ reproduce the results already existing in the literature \cite{Frolov:2006qe,Casteill:2007ct,Belitsky:2007kf,Gromov:2008en,Volin:2008kd}
with
\begin{eqnarray}
      \delta E_{3,{\rm m}}=-\frac{4}{a}\int_a^\infty \frac{dy}{2\pi }\frac{\sqrt{y^2-a^2}}{y^2-1}p'=\frac{a-\left(a^2+1\right)\arccoth \,a}{a\pi}\log\frac{2S}{ag}+\frac{4\arccoth \,a}{a\pi}
\end{eqnarray}
and
\begin{eqnarray}
&\ & \quad \delta E_{1}=-\frac{4}{a}\int_{U_+}\frac{dy}{2\pi}{\,\rm Im\,} \frac{\sqrt{a^2-y^2}-\sqrt{a^2-1}}{y^2-1}\partial_y{\,\rm Im\,} G_0(y)\nn \\ \nn
&\ & \quad =-\frac{
 (a^{2}+1)\arccoth\, a^2+2a^{2}\log (1-a^{-2})+1}{ a \pi}\log\frac{2S}{ag}\\
 &\ &+\frac{1}{a\pi}\(4a \arccot a-4\sqrt{a^2-1}\arccot\sqrt{a^2-1}+2 \log (1-a^{-4})\)
\end{eqnarray}
At $\ell=0$ we get as expected
\begin{eqnarray}
      \delta E_{1}+ \delta E_{3,{\rm m}}=-\frac{3\log 2}{\pi}\log \frac{ 2S}{g} +\frac{6\log 2}{\pi}+1\;.
\end{eqnarray}

\subsection{\label{sec:subseqtionfromY}The wrapping corrections from the Y-system}
As it was shown in \cite{Gromov:2009tq}, exactly the same results which were obtained in the the previous section can be alternatively obtained
directly from the Y-system, without reference to the algebraic curve.
In this subsection we give such an alternative derivation. We find it instructive to identify the origin of different
corrections from the point of view of the Y-system, and this exercise may shed some light on the relation between the two approaches.

According to \cite{Gromov:2009tq} the one-loop wrapping correction to the energy can be computed in terms of the following object
\begin{equation}
\CM_0= \log\frac{(f\D-1)^4(\bar f \D-1)^4}{(\D-1)^4(f \bar f \D-1)^2(f^2 \D-1)(\bar f^2 \D-1)}
\end{equation}
where
\begin{equation}
f(z)=\exp\(-iG(x(z)\)\;, \quad \bar f(z)=\exp\(+iG(1/x(z)\)\;, \quad \D=\exp\(-\frac{J}{2g\sqrt{1-z^2}}\)
\end{equation}
and $G(x)$ is the resolvent
\begin{equation}
G(x)=\frac{1}{g}\sum_{j=1}^S\frac{1}{x-x_j}\frac{x_j^2}{x_j^2-1}\;.
\end{equation}
The expression of the energy at one loop, including the finite-size correction, is
\begin{equation}
\label{enoneloop}
E=\sum_{j=1}^S\frac{x_j^2+1}{x_j^2-1}+\int_{-1}^1\frac{dz}{2\pi}\frac{z}{\sqrt{1-z^2}}\partial_z\CM_0=\sum_{j=1}^S\frac{x_j^2+1}{x_j^2-1}-\int_{-1}^1\frac{dz}{2\pi}\frac{1}{(1-z^2)^{3/2}}\CM_0\;.
\end{equation}
The integration is done in the mirror regime, with $x(z)=z+i\sqrt{1-z^2}$. The second term in (\ref{enoneloop}) is given by the contribution of the virtual particles circulating along the circumference of the system and which scatter with the magnons  with rapidity $x_j$. We are therefore going to call this term the virtual particle contribution. In finite volume, the positions of the Bethe roots $x_j$ are slightly shifted from their infinite volume positions due to their interaction with the virtual particles; we are going to call this effect backreaction. In the one-loop limit, the backreaction can be taken into account \cite{Gromov:2009tq} by adding an extra potential term to the Bethe ansatz equations, which become
\begin{eqnarray}
\label{bckr}
2\pi n &=& p(x+i0)+p(x-i0)+\alpha(x) p'(x)\cot p(x)+\CV(x)\\ \nn
&-&2i\sum_{k=1}^M \int_{-1}^1 dz \(r(x,z)\CM_+-r(1/x,z)\CM_-+u(x,z)\CM_0\)
\end{eqnarray}
with
\begin{equation}
p(x)=\frac{J}{2g}\frac{x}{x^2-1}+G(x)\quad {\rm and}\quad \alpha(x)=\frac 1g \frac{x^2}{x^2-1}\;.
\end{equation}
The effective potential in the second line of (\ref{bckr}) is given in term of the kernels
\begin{equation}
r(x,z)=\frac{x}{x^2-1}\frac{\partial_z}{2\pi g}\frac{1}{x-x(z)}, \qquad u(x,z)=\frac{x}{x^2-1}\frac{\partial_z}{2\pi g}\frac{1}{x^2(z)-1}\;,
\end{equation}
and the functions
\begin{equation}
\CM_+= \log\frac{(f\D-1)^2}{(f^2\D-1)(f\bar f \D-1)}\;, \qquad \CM_-= \log\frac{(\bar f\D-1)^2}{(\bar f^2\D-1)(f\bar f \D-1)}\;.
\end{equation}
By inspection, the imaginary part of the resolvent $G(x)$ in the mirror regime is always negative with ${\rm Im\;}G(x)\sim- \log S$, so that we have
\begin{equation}
 \CM_0\simeq - 4\log({1-\D}) -\D (f^2+\bar f^2+2f\bar f -4f-4\bar f)=- 4\log({1-\D}) -4\D R (R-2)\;,
\end{equation}
where $R=\exp({{\rm Im\;}  G(x)} )\cos({\rm Re\;}  G(x))$. The last term in $\CM_0$  is suppressed by a negative power of $S$. The only region where $R$ can be close to $1$ is $x\simeq 1$, but in this region it is  $\Delta$ which is exponentially suppressed. We conclude that the correction to the energy due to the virtual particles is
\begin{equation}
\delta E_v=4\int_{-1}^1\frac{dz}{2\pi}\frac{1}{(1-z^2)^{3/2}}\log({1-\D})=\CI(2\ell \log S) \;.
\end{equation}
with $\CI(\alpha)$ defined in (\ref{massmodesAC}) and (\ref{expnI}).
It is interesting to note that the virtual particle correction is singular when $\ell \to 0$, and that this divergence will be compensated by the backreaction of the roots.
It is likely that  such a phenomenon happens whenever the endpoint $a$ of the cut approaches the singularity $x=1$. In particular (logarithmic) singularities appear for separate $E_i$ terms in the small $\CS,\CJ$ limit, see section \ref{sec:shortoperatorlimit}. A similar effect is observed when $f_2(\ell)$ (see (\ref{logSexpansion})) is expanded at small $\ell$ \cite{Gromov:2008en}. This partially reflects the complicated analytical structure of the Y-system.

Let us now compute the backreaction term, {\it i.e.} the second line in the BES equation (\ref{bckr}). The contribution from $\CM_\pm$ is vanishing again as a negative
power of $S$. The term containing $\CM_0$ is simply
\begin{eqnarray}
8i \frac{x}{x^2-1}\int_{-1}^1 \frac{dz}{2\pi g} \partial_z\(\frac{1}{x(z)^2-1}\) \log (1-\Delta)=\frac{\;\CI(\alpha)}{g} \frac{x}{x^2-1}\;.
\end{eqnarray}
Let us remind that at the leading order the asymptotic Bethe ansatz equations are written as
\begin{eqnarray}
\label{loeq}
2\pi n= G_0(x+i0)+G_0(x-i0)+2V_0(x)
\end{eqnarray}
with
\begin{eqnarray}
2V_0(x)= \frac{J}{g} \frac{x}{x^2-1}=4\ell \log S \frac{x}{x^2-1}\;
\end{eqnarray}
and $G_0(x)$ a function analytic everywhere except of the cuts  on the intervals $(-\infty,-a)\;\cup\;(a,\infty)$.
We conclude that the only effect of the backreaction at one loop is to renormalize the coefficient of the potential term and therefore to renormalize $\ell$ by a
one-loop quantity
\begin{eqnarray}
\label{shiftell}
\ell \to\tilde \ell= \ell+\frac{\CI(\alpha)}{4g\log S} \simeq \ell -\frac{\pi}{12 \ell g \log^2 S}\;.
\end{eqnarray}
To fix uniquely the solution of the leading order equation (\ref{loeq}) we have to supply the asymptotics at infinity for $G_0(x)$. This is a little bit tricky if we have sent the endpoints of the two cuts $(-b,-a)$ and  $(a,b)$  to infinity first. A straightforward procedure is to solve the equation (\ref{loeq}) for finite $b$ (see \cite{Casteill:2007ct}), imposing that $G_0(x)\sim 1/x$ at $x\to \infty$ and then take the limit of infinite $b$. The result of this procedure would give $G_0(x)=p_{\hat 2}(x)-p_{\tilde 2}(x)$ with $p_{\hat 2}(x),\ p_{\tilde 2}(x)$ from (\ref{eq:p_a}). An alternative is to work directly with $b\to \infty$ and impose the same asymptotics for $G_0(x)$ as in the weak-coupling, one loop limit \cite{Korchemsky:1995be}
\begin{eqnarray}
\label{asinf}
G_0(x)\sim 2i\log (S/gx)\quad {\rm for} \quad x\to \infty-i0\;.
\end{eqnarray}
The solution to the equation (\ref{loeq}) supplemented with this condition at infinity reads
\begin{eqnarray}
G_0(x)&=&\sqrt{a^2-x^2}\oint_{\CC} \frac{dy}{2\pi i}\frac{V_0(y)}{(x-y)\sqrt{a^2-y^2}}-4\arctan\sqrt{\frac{a-x}{a+x}}\\
&=&\frac{J}{2g}\frac{x}{x^2-1}\(\frac{\sqrt{a^2-x^2}}{\sqrt{a^2-1}}-1\)-4\arctan\sqrt{\frac{a-x}{a+x}}\,,
\end{eqnarray}
where in the first line the contour of integration ${\CC}$ encircles the cuts  $(-\infty,-a)\;\cup\;(a,\infty)$ and can be closed at  infinity counterclockwise.
The value of $a$ is fixed by the asymptotics at infinity (\ref{asinf}) and it yields the same condition as (\ref{ella})
\begin{eqnarray}
\label{asycond}
\sqrt{a^2-1}=\frac{J}{4g\log ({2S}/{ag})}=\ell+\CO(1/\log S)\;.
\end{eqnarray}
The anomalous dimension at leading order
is given by
\begin{eqnarray}
E_0&=&- \oint _{\CC}\frac{dx}{2\pi i}\frac{2}{x^2-1} \frac{G_0(x)}{\alpha(x)}=\frac{J}{\sqrt{a^2-1}}\(a-\sqrt{a^2-1}\)-\frac{4g}{a}\nn \\
&=&4g\log S\(\sqrt{1+\ell^2}-\ell\)+\CO((\log S)^0)
\end{eqnarray}
with the integration contour running counterclockwise around $x=0$.
Now we can estimate the one-loop correction from the backreaction due to the shift $\ell \to \ell +\delta \ell$ from equation (\ref{shiftell}),
\begin{eqnarray}
\delta E_b=4g \log S\; \delta \(\sqrt{1+\ell^2}-\ell \)=-\CI(2\ell \log S)+\frac{\ell \CI(2\ell \log S)}{\sqrt{1+\ell^2}}
\end{eqnarray}
The one-loop wrapping corrections are then given by:
\begin{eqnarray}
\label{massmodesY}
\delta E_w=\delta E_v+ \delta E_b=\frac{\ell \CI(2\ell\log S)}{\sqrt{1+\ell^2}}=-\frac{4\pi}{12\log S}+\CO((\ell \log S) \log( \ell \log S))\;
\end{eqnarray}
and they coincide with the contribution of the four massive modes $\delta E_2$ (\ref{massmodesel}).

The wrapping corrections give the $1/\log S$ corrections corresponding to only four of the five bosonic modes. To find the fifth one
we are going to solve the one-loop equation for the resolvent
\begin{eqnarray}
\label{noeq}
0= G_1(x+i0)+G_1(x-i0)+2V_1(x)
\end{eqnarray}
with
\begin{eqnarray}
\frac{2V_1(x)}g=\CV(x)+\alpha(x)\,p_0'\cot p_0\;,
 \qquad p_0(x)=G_0(x)+V_0(x)\label{p0g0v0}\;.
\end{eqnarray}
Here  $\alpha(x)\,p_0'\cot p_0$ is the so-called anomaly term and $\CV(x)$ is the Hernandez-Lopez phase with integral representation \cite{Gromov:2007cd}
\begin{eqnarray}
\CV(x)=\int_{U^+}\frac{dy}{2\pi}\(\frac{\alpha(x)}{x-y}-\frac{\alpha(1/x)}{1/x-y}\)\partial_y(G_0(y)-G_0(1/y))\;,
\end{eqnarray}
where the integral is taken clockwise on the upper half of the unit circle $U_+$.
The solution to the one-loop equation can be again written in an integral form \cite{Casteill:2007ct, Belitsky:2007kf}
\begin{eqnarray}
G_1(x)=\oint_{\CC} \frac{dy}{2\pi i}\frac{V_1(y)}{(x-y)}\frac{\sqrt{a^2-y^2}}{\sqrt{a^2-x^2}}\;.
\end{eqnarray}
The one-loop correction to the energy is given, similarly to the leading order, by
\begin{eqnarray}
E_{{\rm 1,ABA}}=-\oint _{\CC} \frac{dx}{2\pi i}\frac{2G_1(x)}{x^2}=\frac{2}{a}\oint_{\CC} \frac{dy}{2\pi i}\frac{\sqrt{a^2-y^2}}{y^2}V_1(y)\;.
\end{eqnarray}
Substituting the value of the potential $V_1(x)$ we retrieve the contributions (\ref{mainc}) from the algebraic curve computation
\begin{eqnarray}
E_{{\rm 1,ABA}}&=&-\frac{4}{a}\int_a^\infty \frac{dy}{2\pi }\frac{\sqrt{y^2-a^2}}{y^2-1}p'\coth p-\frac{4}{a}\int_{U_+}\frac{dy}{2\pi}{\,\rm Im\,} \frac{\sqrt{a^2-y^2}-\sqrt{a^2-1}}{y^2-1}\partial_y{\,\rm Im\,} G_0(y)\nn \\
&=&\delta E_1+\delta E_3\;.
\end{eqnarray}
This result confirms that the asymptotic Bethe ansatz contribution is captured by $\delta E_1+\delta E_3$ and that the wrapping corrections are reproduced by
$\delta E_2$.

\bigskip
\leftline{\bf \large Conclusion}

\noindent We have computed one-loop correction to the energy of the folded string in the limit of large angular momentum in $AdS_5$
and on the sphere, by using the algebraic curve method and alternatively the Y-system method. By extrapolating the results at small angular momenta, we obtain predictions for  the first few coefficients in the strong coupling expansion. The result confirms the numerical prediction from the Y-system for the Konishi operator \cite{Gromov:2009zb} and more recent numerical results for similar operators \cite{Gromov:2011xx}.
In the limit of large Lorentz spin and small R-charge, we have computed the corrections to the logarithmic scaling and in particular we have reproduced
the $1/\log S$ corrections obtained in \cite{SchaferNameki:2005is,Beccaria:2010ry,Giombi:2010zi}.

\bigskip
\leftline{\bf \large Acknowledgments}

\noindent We would like to thank Benjamin Basso, Vladimir Kazakov,
Gregory Korchemsky, Ivan Kostov, Sebastien Leurent, Radu Roiban, and Arkady Tseytlin for
interesting discussions. The work of D.V. was
supported by the US Department of Energy under contracts
DE-FG02-201390ER40577. I.S. was partially supported by the grant RFFI
11-01-00570. N.G. was partially supported by the grant RFFI
06-02-16786.\\

\bigskip \leftline{\bf Note Added}
\noindent While we were preparing these
results for publication we became aware that the authors of
\cite{Roiban:2009aa} also found a way to correct their previous result
concerning the Konishi dimension.  This will be published in
\cite{Roiban:2011fe}. A different approach to compute analytically the
Konishi dimension in the pure spinor formalism was developed in
\cite{Vallilo:2011fj}.
\\ \leftline{\bf Note Added for v4}
We restrict the discussion of short operators to the case $n=1$. The case $n>1$ may require a further consideration. We thank Matteo Beccaria for discussing this point. This restriction does not affect our main conclusion about short operators - the agreement with available numerical solutions of TBA.

\appendix
\section{Simplified form of the one-loop integrals}
\la{AppA}

Using symmetry relations \eqref{eq:quasimomenta_symmetry_A},
\eqref{eq:quasimomenta_symmetry_S} \eqref{eq:frequencies}, one can
rewrite sums $\delta E_1$ and $\delta E_2$ through the functions
$p_{\hat{2}}(x)$, $p_{\tilde{2}}(x)$, $\Omega^{\hat{2}\hat{3}}(x)$,
$\Omega^{\tilde{2}\tilde{3}}(x)$ defined above.

Let us consider for example the set of polarizations which belongs to
the $S^5$. As we already have seen, all the frequencies are equal to
$\Omega^{\tilde{2}\tilde{3}}(x)$. So our sum in $\delta E_2$
simplifies drastically and gives
\begin{equation}
  \label{eq:sphere_sum}
  \delta E_2^{S^5} = 4 \int\limits_{0}^{1} \frac{dz}{\pi} \Im
  \left[ \Omega^{\tilde{2}\tilde{3}}(z) \pt_z \log \left(1-e^{-2ip_{\tilde{2}}(z)} \right) \right],
\end{equation}
which exactly coincides with the expression \eqref{massmodesAC}.

One can easily show that all the contributions in $\delta E_{1,2}$ can
be very naturally summed up in a pretty nice way:

\begin{eqnarray}
  \label{eq:delta_E_final}
  \nn
  \delta E_1&=& 2\int\limits_0^1
  \frac{dz}{\pi}\Im(p_{\hat{2}}-p_{\tilde{2}})\partial_z \Im (\Omega^{\hat{2}\hat{3}}-\Omega^{\tilde{2}\tilde{3}}),\\
  \delta
  E_2 &=& 2\int\limits_0^1\frac{dz}{\pi} \Im \left( \partial_z\Omega^{\tilde{2}
      \tilde{3}} \log \frac{(1-e^{-i p_{\tilde{2}}-i \bar{p}_{\hat{2}}})(1-e^{-i
        p_{\tilde{2}}+ip_{\hat{2}}})}{(1-e^{-2ip_{\tilde{2}}})^2} - \right.\\
  \nn
  &-&\left. \partial_z\Omega^{\hat{2}\hat{3}} \log\frac{(1-e^{-2ip_{\hat{2}}})(1-e^{-i
        p_{\hat{2}}+i\bar{p}_{\hat{2}}})}{(1-e^{-ip_{\hat{2}}-i p_{\tilde{2}}})^2} \right).
\end{eqnarray}

Here for shortness we denote $\bar{p}(z) = p(1/z)$.

\bibliographystyle{bibstyle}
\bibliography{konishi}

\providecommand{\href}[2]{#2}\begingroup\raggedright\begin{thebibliography}{10}

\bibitem{Maldacena:1997re}
J.~M. Maldacena, ``{The large N limit of superconformal field theories and
  supergravity}'', \href{http://dx.doi.org/10.1023/A:1026654312961}{{\em Adv.
  Theor. Math. Phys.} {\bf 2} (1998)  231--252},
\href{http://arXiv.org/abs/hep-th/9711200}{\texttt{arXiv:hep-th/9711200}}.

\bibitem{Witten:1998qj}
E.~Witten, ``{Anti-de Sitter space and holography}'', {\em Adv. Theor. Math.
  Phys.} {\bf 2} (1998)  253--291,
\href{http://arXiv.org/abs/hep-th/9802150}{\texttt{arXiv:hep-th/9802150}}.

\bibitem{Gubser:1998bc}
S.~S. Gubser, I.~R. Klebanov, and A.~M. Polyakov, ``{Gauge theory correlators
  from non-critical string theory}'',
  \href{http://dx.doi.org/10.1016/S0370-2693(98)00377-3}{{\em Phys. Lett.} {\bf
  B428} (1998)  105--114},
\href{http://arXiv.org/abs/hep-th/9802109}{\texttt{arXiv:hep-th/9802109}}.

\bibitem{Lipatov:1993yb}
L.~N. Lipatov, ``{High-energy asymptotics of multicolor QCD and exactly
  solvable lattice models}'',
\href{http://arXiv.org/abs/hep-th/9311037}{\texttt{arXiv:hep-th/9311037}}.

\bibitem{Faddeev:1994zg}
L.~D. Faddeev and G.~P. Korchemsky, ``{High-energy QCD as a completely
  integrable model}'',
  \href{http://dx.doi.org/10.1016/0370-2693(94)01363-H}{{\em Phys. Lett.} {\bf
  B342} (1995)  311--322},
\href{http://arXiv.org/abs/hep-th/9404173}{\texttt{arXiv:hep-th/9404173}}.

\bibitem{Minahan:2002ve}
J.~A. Minahan and K.~Zarembo, ``{The Bethe-ansatz for N = 4 super
  Yang-Mills}'', {\em JHEP} {\bf 03} (2003)  013,
\href{http://arXiv.org/abs/hep-th/0212208}{\texttt{arXiv:hep-th/0212208}}.

\bibitem{Beisert:2003tq}
N.~Beisert, C.~Kristjansen, and M.~Staudacher, ``{The dilatation operator of N
  = 4 super Yang-Mills theory}'',
  \href{http://dx.doi.org/10.1016/S0550-3213(03)00406-1}{{\em Nucl. Phys.} {\bf
  B664} (2003)  131--184},
\href{http://arXiv.org/abs/hep-th/0303060}{\texttt{arXiv:hep-th/0303060}}.

\bibitem{Bena:2003wd}
I.~Bena, J.~Polchinski, and R.~Roiban, ``{Hidden symmetries of the $AdS_5
  \times S^5$ superstring}'',
  \href{http://dx.doi.org/10.1103/PhysRevD.69.046002}{{\em Phys. Rev.} {\bf
  D69} (2004)  046002},
\href{http://arXiv.org/abs/hep-th/0305116}{\texttt{arXiv:hep-th/0305116}}.

\bibitem{Kazakov:2004qf}
V.~A. Kazakov, A.~Marshakov, J.~A. Minahan, and K.~Zarembo, ``{Classical /
  quantum integrability in AdS/CFT}'',
  \href{http://dx.doi.org/10.1088/1126-6708/2004/05/024}{{\em JHEP} {\bf 05}
  (2004)  024},
\href{http://arXiv.org/abs/hep-th/0402207}{\texttt{arXiv:hep-th/0402207}}.

\bibitem{Beisert:2010jr}
N.~Beisert {\em et al.}, ``{Review of AdS/CFT Integrability: An Overview}'',
\href{http://arXiv.org/abs/1012.3982}{\texttt{arXiv:1012.3982}}.

\bibitem{Beisert:2003yb}
N.~Beisert and M.~Staudacher, ``{The N=4 SYM Integrable Super Spin Chain}'',
  \href{http://dx.doi.org/10.1016/j.nuclphysb.2003.08.015}{{\em Nucl. Phys.}
  {\bf B670} (2003)  439--463},
\href{http://arXiv.org/abs/hep-th/0307042}{\texttt{arXiv:hep-th/0307042}}.

\bibitem{Beisert:2005fw}
N.~Beisert and M.~Staudacher, ``{Long-range psu(2,2|4) Bethe Ansatze for gauge
  theory and strings}'',
  \href{http://dx.doi.org/10.1016/j.nuclphysb.2005.06.038}{{\em Nucl.Phys.}
  {\bf B727} (2005)  1--62},
  \href{http://arXiv.org/abs/hep-th/0504190}{\texttt{arXiv:hep-th/0504190}}. In
  honor of Hans Bethe.

\bibitem{Beisert:2006ez}
N.~Beisert, B.~Eden, and M.~Staudacher, ``{Transcendentality and crossing}'',
  \href{http://dx.doi.org/10.1088/1742-5468/2007/01/P01021}{{\em J. Stat.
  Mech.} {\bf 0701} (2007)  P021},
\href{http://arXiv.org/abs/hep-th/0610251}{\texttt{arXiv:hep-th/0610251}}.

\bibitem{Gromov:2009tv}
N.~Gromov, V.~Kazakov, and P.~Vieira, ``{Exact Spectrum of Anomalous Dimensions
  of Planar N=4 Supersymmetric Yang-Mills Theory}'',
  \href{http://dx.doi.org/10.1103/PhysRevLett.103.131601}{{\em Phys. Rev.
  Lett.} {\bf 103} (2009)  131601},
\href{http://arXiv.org/abs/0901.3753}{\texttt{arXiv:0901.3753}}.

\bibitem{Bombardelli:2009ns}
D.~Bombardelli, D.~Fioravanti, and R.~Tateo, ``{Thermodynamic Bethe Ansatz for
  planar AdS/CFT: A Proposal}'',
  \href{http://dx.doi.org/10.1088/1751-8113/42/37/375401,
  10.1088/1751-8113/42/37/375401}{{\em J.Phys.A} {\bf A42} (2009)  375401},
  \href{http://arXiv.org/abs/0902.3930}{\texttt{arXiv:0902.3930}}.

\bibitem{Gromov:2009bc}
N.~Gromov, V.~Kazakov, A.~Kozak, and P.~Vieira, ``{Exact Spectrum of Anomalous
  Dimensions of Planar N = 4 Supersymmetric Yang-Mills Theory: TBA and excited
  states}'', \href{http://dx.doi.org/10.1007/s11005-010-0374-8}{{\em Lett.
  Math. Phys.} {\bf 91} (2010)  265--287},
\href{http://arXiv.org/abs/0902.4458}{\texttt{arXiv:0902.4458}}.

\bibitem{Arutyunov:2009ur}
G.~Arutyunov and S.~Frolov, ``{Thermodynamic Bethe Ansatz for the $AdS_5 \times
  S^5$ Mirror Model}'',
  \href{http://dx.doi.org/10.1088/1126-6708/2009/05/068}{{\em JHEP} {\bf 0905}
  (2009)  068},
  \href{http://arXiv.org/abs/0903.0141}{\texttt{arXiv:0903.0141}}.

\bibitem{Cavaglia:2010nm}
A.~Cavaglia, D.~Fioravanti, and R.~Tateo, ``{Extended Y-system for the
  $AdS_5/CFT_4$ correspondence}'',
  \href{http://dx.doi.org/10.1016/j.nuclphysb.2010.09.015}{{\em Nucl. Phys.}
  {\bf B843} (2011)  302--343},
\href{http://arXiv.org/abs/1005.3016}{\texttt{arXiv:1005.3016}}.

\bibitem{Gromov:2009tq}
N.~Gromov, ``{Y-system and Quasi-Classical Strings}'',
  \href{http://dx.doi.org/10.1007/JHEP01(2010)112}{{\em JHEP} {\bf 01} (2010)
  112},
\href{http://arXiv.org/abs/0910.3608}{\texttt{arXiv:0910.3608}}.

\bibitem{Gromov:2010vb}
N.~Gromov, V.~Kazakov, and Z.~Tsuboi, ``{PSU(2,2|4) Character of Quasiclassical
  AdS/CFT}'', \href{http://dx.doi.org/10.1007/JHEP07(2010)097}{{\em JHEP} {\bf
  07} (2010)  097},
\href{http://arXiv.org/abs/1002.3981}{\texttt{arXiv:1002.3981}}.

\bibitem{Fiamberti:2007rj}
F.~Fiamberti, A.~Santambrogio, C.~Sieg, and D.~Zanon, ``{Wrapping at four loops
  in N=4 SYM}'', \href{http://dx.doi.org/10.1016/j.physletb.2008.06.061}{{\em
  Phys. Lett.} {\bf B666} (2008)  100--105},
\href{http://arXiv.org/abs/0712.3522}{\texttt{arXiv:0712.3522}}.

\bibitem{Velizhanin:2008jd}
V.~N. Velizhanin, ``{The four-loop anomalous dimension of the Konishi operator
  in N=4 supersymmetric Yang-Mills theory}'',
  \href{http://dx.doi.org/10.1134/S0021364009010020}{{\em JETP Lett.} {\bf 89}
  (2009)  6--9},
\href{http://arXiv.org/abs/0808.3832}{\texttt{arXiv:0808.3832}}.

\bibitem{Arutyunov:2010gb}
G.~Arutyunov, S.~Frolov, and R.~Suzuki, ``{Five-loop Konishi from the Mirror
  TBA}'', \href{http://dx.doi.org/10.1007/JHEP04(2010)069}{{\em JHEP} {\bf 04}
  (2010)  069},
\href{http://arXiv.org/abs/1002.1711}{\texttt{arXiv:1002.1711}}.

\bibitem{Balog:2010xa}
J.~Balog and A.~Hegedus, ``{5-loop Konishi from linearized TBA and the XXX
  magnet}'', \href{http://dx.doi.org/10.1007/JHEP06(2010)080}{{\em JHEP} {\bf
  06} (2010)  080},
\href{http://arXiv.org/abs/1002.4142}{\texttt{arXiv:1002.4142}}.

\bibitem{Balog:2010vf}
J.~Balog and A.~Hegedus, ``{The Bajnok-Janik formula and wrapping
  corrections}'', \href{http://dx.doi.org/10.1007/JHEP09(2010)107}{{\em JHEP}
  {\bf 1009} (2010)  107},
  \href{http://arXiv.org/abs/1003.4303}{\texttt{arXiv:1003.4303}}.

\bibitem{Bajnok:2008bm}
Z.~Bajnok and R.~A. Janik, ``{Four-loop perturbative Konishi from strings and
  finite size effects for multiparticle states}'',
  \href{http://dx.doi.org/10.1016/j.nuclphysb.2008.08.020}{{\em Nucl. Phys.}
  {\bf B807} (2009)  625--650},
\href{http://arXiv.org/abs/0807.0399}{\texttt{arXiv:0807.0399}}.

\bibitem{Bajnok:2009vm}
Z.~Bajnok, A.~Hegedus, R.~A. Janik, and T.~Lukowski, ``{Five loop Konishi from
  AdS/CFT}'', \href{http://dx.doi.org/10.1016/j.nuclphysb.2009.10.015}{{\em
  Nucl. Phys.} {\bf B827} (2010)  426--456},
\href{http://arXiv.org/abs/0906.4062}{\texttt{arXiv:0906.4062}}.

\bibitem{Kotikov:2007cy}
A.~V. Kotikov, L.~N. Lipatov, A.~Rej, M.~Staudacher, and V.~N. Velizhanin,
  ``{Dressing and Wrapping}'',
  \href{http://dx.doi.org/10.1088/1742-5468/2007/10/P10003}{{\em J. Stat.
  Mech.} {\bf 0710} (2007)  P10003},
\href{http://arXiv.org/abs/0704.3586}{\texttt{arXiv:0704.3586}}.

\bibitem{Bajnok:2008qj}
Z.~Bajnok, R.~A. Janik, and T.~Lukowski, ``{Four loop twist two, BFKL, wrapping
  and strings}'', \href{http://dx.doi.org/10.1016/j.nuclphysb.2009.02.005}{{\em
  Nucl. Phys.} {\bf B816} (2009)  376--398},
\href{http://arXiv.org/abs/0811.4448}{\texttt{arXiv:0811.4448}}.

\bibitem{Lukowski:2009ce}
T.~Lukowski, A.~Rej, and V.~N. Velizhanin, ``{Five-Loop Anomalous Dimension of
  Twist-Two Operators}'',
  \href{http://dx.doi.org/10.1016/j.nuclphysb.2010.01.008}{{\em Nucl. Phys.}
  {\bf B831} (2010)  105--132},
\href{http://arXiv.org/abs/0912.1624}{\texttt{arXiv:0912.1624}}.

\bibitem{Kotikov:2002ab}
A.~Kotikov and L.~Lipatov, ``{DGLAP and BFKL equations in the N=4
  supersymmetric gauge theory}'', {\em Nucl.Phys.} {\bf B661} (2003)  19--61,
  \href{http://arXiv.org/abs/hep-ph/0208220}{\texttt{arXiv:hep-ph/0208220}}.

\bibitem{Gromov:2009zb}
N.~Gromov, V.~Kazakov, and P.~Vieira, ``{Exact Spectrum of Planar ${\cal N}=4$
  Supersymmetric Yang- Mills Theory: Konishi Dimension at Any Coupling}'',
  \href{http://dx.doi.org/10.1103/PhysRevLett.104.211601}{{\em Phys. Rev.
  Lett.} {\bf 104} (2010)  211601},
\href{http://arXiv.org/abs/0906.4240}{\texttt{arXiv:0906.4240}}.

\bibitem{Frolov:2010wt}
S.~Frolov, ``{Konishi operator at intermediate coupling}'',
  \href{http://dx.doi.org/10.1088/1751-8113/44/6/065401}{{\em J. Phys.} {\bf
  A44} (2011)  065401},
\href{http://arXiv.org/abs/1006.5032}{\texttt{arXiv:1006.5032}}.

\bibitem{Gubser:2002tv}
S.~S. Gubser, I.~R. Klebanov, and A.~M. Polyakov, ``{A semi-classical limit of
  the gauge/string correspondence}'',
  \href{http://dx.doi.org/10.1016/S0550-3213(02)00373-5}{{\em Nucl. Phys.} {\bf
  B636} (2002)  99--114},
\href{http://arXiv.org/abs/hep-th/0204051}{\texttt{arXiv:hep-th/0204051}}.

\bibitem{Passerini:2010xc}
F.~Passerini, J.~Plefka, G.~W. Semenoff, and D.~Young, ``{On the Spectrum of
  the $AdS_5 \times S^5$ String at large lambda}'',
  \href{http://dx.doi.org/10.1007/JHEP03(2011)046}{{\em JHEP} {\bf 1103} (2011)
   046}, \href{http://arXiv.org/abs/1012.4471}{\texttt{arXiv:1012.4471}}.

\bibitem{Roiban:2009aa}
R.~Roiban and A.~A. Tseytlin, ``{Quantum strings in $AdS_5 \times S^5$:
  strong-coupling corrections to dimension of Konishi operator}'',
  \href{http://dx.doi.org/10.1088/1126-6708/2009/11/013}{{\em JHEP} {\bf 11}
  (2009)  013},
\href{http://arXiv.org/abs/0906.4294}{\texttt{arXiv:0906.4294}}.

\bibitem{Arutyunov:2009ax}
G.~Arutyunov, S.~Frolov, and R.~Suzuki, ``{Exploring the mirror TBA}'',
  \href{http://dx.doi.org/10.1007/JHEP05(2010)031}{{\em JHEP} {\bf 05} (2010)
  031},
\href{http://arXiv.org/abs/0911.2224}{\texttt{arXiv:0911.2224}}.

\bibitem{SchaferNameki:2010jy}
S.~Schafer-Nameki, ``{Review of AdS/CFT Integrability, Chapter II.4: The
  Spectral Curve}'',
\href{http://arXiv.org/abs/1012.3989}{\texttt{arXiv:1012.3989}}.

\bibitem{Gromov:2011xx}
N.~Gromov, V.~Kazakov, S.~Leurent, and D.~Volin. {Work in progress}.

\bibitem{Frolov:2006qe}
S.~Frolov, A.~Tirziu, and A.~A. Tseytlin, ``{Logarithmic corrections to higher
  twist scaling at strong coupling from AdS/CFT}'',
  \href{http://dx.doi.org/10.1016/j.nuclphysb.2006.12.013}{{\em Nucl. Phys.}
  {\bf B766} (2007)  232--245},
\href{http://arXiv.org/abs/hep-th/0611269}{\texttt{arXiv:hep-th/0611269}}.

\bibitem{Casteill:2007ct}
P.~Y. Casteill and C.~Kristjansen, ``{The Strong Coupling Limit of the Scaling
  Function from the Quantum String Bethe Ansatz}'',
  \href{http://dx.doi.org/10.1016/j.nuclphysb.2007.06.011}{{\em Nucl. Phys.}
  {\bf B785} (2007)  1--18},
\href{http://arXiv.org/abs/0705.0890}{\texttt{arXiv:0705.0890}}.

\bibitem{Belitsky:2007kf}
A.~V. Belitsky, ``{Strong coupling expansion of Baxter equation in N=4 SYM}'',
  \href{http://dx.doi.org/10.1016/j.physletb.2007.11.023}{{\em Phys. Lett.}
  {\bf B659} (2008)  732--740},
\href{http://arXiv.org/abs/0710.2294}{\texttt{arXiv:0710.2294}}.

\bibitem{Gromov:2008en}
N.~Gromov, ``{Generalized Scaling Function at Strong Coupling}'',
  \href{http://dx.doi.org/10.1088/1126-6708/2008/11/085}{{\em JHEP} {\bf 11}
  (2008)  085},
\href{http://arXiv.org/abs/0805.4615}{\texttt{arXiv:0805.4615}}.

\bibitem{Gross:1974cs}
D.~J. Gross and F.~Wilczek, ``{Asymptotically free gauge theories. 2}'',
\href{http://dx.doi.org/10.1103/PhysRevD.9.980}{{\em Phys. Rev.} {\bf D9}
  (1974)  980--993}.

\bibitem{Korchemsky:1988si}
G.~P. Korchemsky, ``{Asymptotics of the Altarelli-Parisi-Lipatov Evolution
  Kernels of Parton Distributions}'',
\href{http://dx.doi.org/10.1142/S0217732389001453}{{\em Mod. Phys. Lett.} {\bf
  A4} (1989)  1257--1276}.

\bibitem{Korchemsky:1992xv}
G.~P. Korchemsky and G.~Marchesini, ``{Structure function for large x and
  renormalization of Wilson loop}'',
  \href{http://dx.doi.org/10.1016/0550-3213(93)90167-N}{{\em Nucl. Phys.} {\bf
  B406} (1993)  225--258},
\href{http://arXiv.org/abs/hep-ph/9210281}{\texttt{arXiv:hep-ph/9210281}}.

\bibitem{Belitsky:2006en}
A.~V. Belitsky, A.~S. Gorsky, and G.~P. Korchemsky, ``{Logarithmic scaling in
  gauge / string correspondence}'',
  \href{http://dx.doi.org/10.1016/j.nuclphysb.2006.04.030}{{\em Nucl. Phys.}
  {\bf B748} (2006)  24--59},
\href{http://arXiv.org/abs/hep-th/0601112}{\texttt{arXiv:hep-th/0601112}}.

\bibitem{Volin:2008kd}
D.~Volin, ``{The 2-loop generalized scaling function from the BES/FRS
  equation}'',
\href{http://arXiv.org/abs/0812.4407}{\texttt{arXiv:0812.4407}}.

\bibitem{Giombi:2010fa}
S.~Giombi, R.~Ricci, R.~Roiban, A.~A. Tseytlin, and C.~Vergu, ``{Generalized
  scaling function from light-cone gauge $AdS_5 \times S^5$ superstring}'',
  \href{http://dx.doi.org/10.1007/JHEP06(2010)060}{{\em JHEP} {\bf 06} (2010)
  060},
\href{http://arXiv.org/abs/1002.0018}{\texttt{arXiv:1002.0018}}.

\bibitem{Kotikov:2006ts}
A.~V. Kotikov and L.~N. Lipatov, ``{On the highest transcendentality in N = 4
  SUSY}'', \href{http://dx.doi.org/10.1016/j.nuclphysb.2007.01.020}{{\em Nucl.
  Phys.} {\bf B769} (2007)  217--255},
\href{http://arXiv.org/abs/hep-th/0611204}{\texttt{arXiv:hep-th/0611204}}.

\bibitem{Benna:2006nd}
M.~K. Benna, S.~Benvenuti, I.~R. Klebanov, and A.~Scardicchio, ``{A test of the
  AdS/CFT correspondence using high-spin operators}'',
  \href{http://dx.doi.org/10.1103/PhysRevLett.98.131603}{{\em Phys. Rev. Lett.}
  {\bf 98} (2007)  131603},
\href{http://arXiv.org/abs/hep-th/0611135}{\texttt{arXiv:hep-th/0611135}}.

\bibitem{Alday:2007qf}
L.~F. Alday, G.~Arutyunov, M.~K. Benna, B.~Eden, and I.~R. Klebanov, ``{On the
  strong coupling scaling dimension of high spin operators}'',
  \href{http://dx.doi.org/10.1088/1126-6708/2007/04/082}{{\em JHEP} {\bf 04}
  (2007)  082},
\href{http://arXiv.org/abs/hep-th/0702028}{\texttt{arXiv:hep-th/0702028}}.

\bibitem{Kostov:2007kx}
I.~Kostov, D.~Serban, and D.~Volin, ``{Strong coupling limit of Bethe ansatz
  equations}'', \href{http://dx.doi.org/10.1016/j.nuclphysb.2007.06.017}{{\em
  Nucl. Phys.} {\bf B789} (2008)  413--451},
\href{http://arXiv.org/abs/hep-th/0703031}{\texttt{arXiv:hep-th/0703031}}.

\bibitem{Beccaria:2007tk}
M.~Beccaria, G.~F. De~Angelis, and V.~Forini, ``{The scaling function at strong
  coupling from the quantum string Bethe equations}'',
  \href{http://dx.doi.org/10.1088/1126-6708/2007/04/066}{{\em JHEP} {\bf 04}
  (2007)  066},
\href{http://arXiv.org/abs/hep-th/0703131}{\texttt{arXiv:hep-th/0703131}}.

\bibitem{Basso:2007wd}
B.~Basso, G.~P. Korchemsky, and J.~Kotanski, ``{Cusp anomalous dimension in
  maximally supersymmetric Yang- Mills theory at strong coupling}'',
  \href{http://dx.doi.org/10.1103/PhysRevLett.100.091601}{{\em Phys. Rev.
  Lett.} {\bf 100} (2008)  091601},
\href{http://arXiv.org/abs/0708.3933}{\texttt{arXiv:0708.3933}}.

\bibitem{Kostov:2008ax}
I.~Kostov, D.~Serban, and D.~Volin, ``{Functional BES equation}'',
  \href{http://dx.doi.org/10.1088/1126-6708/2008/08/101}{{\em JHEP} {\bf 08}
  (2008)  101},
\href{http://arXiv.org/abs/0801.2542}{\texttt{arXiv:0801.2542}}.

\bibitem{Eden:2006rx}
B.~Eden and M.~Staudacher, ``{Integrability and transcendentality}'',
  \href{http://dx.doi.org/10.1088/1742-5468/2006/11/P11014}{{\em J. Stat.
  Mech.} {\bf 0611} (2006)  P11014},
\href{http://arXiv.org/abs/hep-th/0603157}{\texttt{arXiv:hep-th/0603157}}.

\bibitem{Freyhult:2010kc}
L.~Freyhult, ``{Review of AdS/CFT Integrability, Chapter III.4: Twist states
  and the cusp anomalous dimension}'',
\href{http://arXiv.org/abs/1012.3993}{\texttt{arXiv:1012.3993}}.

\bibitem{Freyhult:2009my}
L.~Freyhult and S.~Zieme, ``{The virtual scaling function of AdS/CFT}'',
  \href{http://dx.doi.org/10.1103/PhysRevD.79.105009}{{\em Phys. Rev.} {\bf
  D79} (2009)  105009},
\href{http://arXiv.org/abs/0901.2749}{\texttt{arXiv:0901.2749}}.

\bibitem{Fioravanti:2009xt}
D.~Fioravanti, P.~Grinza, and M.~Rossi, ``{Beyond cusp anomalous dimension from
  integrability}'',
  \href{http://dx.doi.org/10.1016/j.physletb.2009.03.053}{{\em Phys.Lett.} {\bf
  B675} (2009)  137--144},
  \href{http://arXiv.org/abs/0901.3161}{\texttt{arXiv:0901.3161}}.

\bibitem{SchaferNameki:2005is}
S.~Schafer-Nameki and M.~Zamaklar, ``{Stringy sums and corrections to the
  quantum string Bethe ansatz}'',
  \href{http://dx.doi.org/10.1088/1126-6708/2005/10/044}{{\em JHEP} {\bf 10}
  (2005)  044},
\href{http://arXiv.org/abs/hep-th/0509096}{\texttt{arXiv:hep-th/0509096}}.

\bibitem{Beccaria:2010ry}
M.~Beccaria, G.~V. Dunne, V.~Forini, M.~Pawellek, and A.~A. Tseytlin, ``{Exact
  computation of one-loop correction to energy of spinning folded string in
  $AdS_5 \times S^5$}'',
  \href{http://dx.doi.org/10.1088/1751-8113/43/16/165402}{{\em J. Phys. A} {\bf
  43} (2010)  165402},
\href{http://arXiv.org/abs/1001.4018}{\texttt{arXiv:1001.4018}}.

\bibitem{Giombi:2010zi}
S.~Giombi, R.~Ricci, R.~Roiban, and A.~A. Tseytlin, ``{Two-loop $AdS_5 \times
  S^5$ superstring: testing asymptotic Bethe ansatz and finite size
  corrections}'', \href{http://dx.doi.org/10.1088/1751-8113/44/4/045402}{{\em
  J. Phys.} {\bf A44} (2011)  045402},
\href{http://arXiv.org/abs/1010.4594}{\texttt{arXiv:1010.4594}}.

\bibitem{Basso:2006nk}
B.~Basso and G.~P. Korchemsky, ``{Anomalous dimensions of high-spin operators
  beyond the leading order}'',
  \href{http://dx.doi.org/10.1016/j.nuclphysb.2007.03.044}{{\em Nucl. Phys.}
  {\bf B775} (2007)  1--30},
\href{http://arXiv.org/abs/hep-th/0612247}{\texttt{arXiv:hep-th/0612247}}.

\bibitem{Fioravanti:2009ei}
D.~Fioravanti, P.~Grinza, and M.~Rossi, ``{On the logarithmic powers of $sl(2)$
  SYM$_4$}'', \href{http://dx.doi.org/10.1016/j.physletb.2009.12.057}{{\em
  Phys. Lett.} {\bf B684} (2010)  52--60},
\href{http://arXiv.org/abs/0911.2425}{\texttt{arXiv:0911.2425}}.

\bibitem{Alday:2007mf}
L.~F. Alday and J.~M. Maldacena, ``{Comments on operators with large spin}'',
  \href{http://dx.doi.org/10.1088/1126-6708/2007/11/019}{{\em JHEP} {\bf 11}
  (2007)  019},
\href{http://arXiv.org/abs/0708.0672}{\texttt{arXiv:0708.0672}}.

\bibitem{Basso:2008tx}
B.~Basso and G.~P. Korchemsky, ``{Embedding nonlinear O(6) sigma model into N=4
  super-Yang- Mills theory}'',
  \href{http://dx.doi.org/10.1016/j.nuclphysb.2008.07.007}{{\em Nucl. Phys.}
  {\bf B807} (2009)  397--423},
\href{http://arXiv.org/abs/0805.4194}{\texttt{arXiv:0805.4194}}.

\bibitem{Fioravanti:2008rv}
D.~Fioravanti, P.~Grinza, and M.~Rossi, ``{Strong coupling for planar N=4 SYM
  theory: An All-order result}'',
  \href{http://dx.doi.org/10.1016/j.nuclphysb.2008.10.018}{{\em Nucl.Phys.}
  {\bf B810} (2009)  563--574},
  \href{http://arXiv.org/abs/0804.2893}{\texttt{arXiv:0804.2893}}.

\bibitem{Basso:2011xx}
B.~Basso. {To appear}.

\bibitem{Gromov:2007aq}
N.~Gromov and P.~Vieira, ``{The AdS(5) x S**5 superstring quantum spectrum from
  the algebraic curve}'',
  \href{http://dx.doi.org/10.1016/j.nuclphysb.2007.07.032}{{\em Nucl. Phys.}
  {\bf B789} (2008)  175--208},
\href{http://arXiv.org/abs/hep-th/0703191}{\texttt{arXiv:hep-th/0703191}}.

\bibitem{Mikhaylov:2010ib}
V.~Mikhaylov, ``{On the Fermionic Frequencies of Circular Strings}'',
  \href{http://dx.doi.org/10.1088/1751-8113/43/33/335401}{{\em J. Phys.} {\bf
  A43} (2010)  335401},
\href{http://arXiv.org/abs/1002.1831}{\texttt{arXiv:1002.1831}}.

\bibitem{Beisert:2003ea}
N.~Beisert, S.~Frolov, M.~Staudacher, and A.~A. Tseytlin, ``{Precision
  spectroscopy of AdS/CFT}'', {\em JHEP} {\bf 10} (2003)  037,
\href{http://arXiv.org/abs/hep-th/0308117}{\texttt{arXiv:hep-th/0308117}}.

\bibitem{Kazakov:2004nh}
V.~A. Kazakov and K.~Zarembo, ``{Classical / quantum integrability in
  non-compact sector of AdS/CFT}'',
  \href{http://dx.doi.org/10.1088/1126-6708/2004/10/060}{{\em JHEP} {\bf 10}
  (2004)  060},
\href{http://arXiv.org/abs/hep-th/0410105}{\texttt{arXiv:hep-th/0410105}}.

\bibitem{Beisert:2005bm}
N.~Beisert, V.~A. Kazakov, K.~Sakai, and K.~Zarembo, ``{The algebraic curve of
  classical superstrings on $AdS_5 \times S^5$}'',
  \href{http://dx.doi.org/10.1007/s00220-006-1529-4}{{\em Commun. Math. Phys.}
  {\bf 263} (2006)  659--710},
\href{http://arXiv.org/abs/hep-th/0502226}{\texttt{arXiv:hep-th/0502226}}.

\bibitem{Gromov:2008ec}
N.~Gromov, S.~Schafer-Nameki, and P.~Vieira, ``{Efficient precision
  quantization in AdS/CFT}'',
  \href{http://dx.doi.org/10.1088/1126-6708/2008/12/013}{{\em JHEP} {\bf 12}
  (2008)  013},
\href{http://arXiv.org/abs/0807.4752}{\texttt{arXiv:0807.4752}}.

\bibitem{Gromov:2007cd}
N.~Gromov and P.~Vieira, ``{Constructing the AdS/CFT dressing factor}'',
  \href{http://dx.doi.org/10.1016/j.nuclphysb.2007.08.019}{{\em Nucl. Phys.}
  {\bf B790} (2008)  72--88},
\href{http://arXiv.org/abs/hep-th/0703266}{\texttt{arXiv:hep-th/0703266}}.

\bibitem{Rej:2009dk}
A.~Rej and F.~Spill, ``{Konishi at strong coupling from ABE}'',
  \href{http://dx.doi.org/10.1088/1751-8113/42/44/442003}{{\em J. Phys.} {\bf
  A42} (2009)  442003},
\href{http://arXiv.org/abs/0907.1919}{\texttt{arXiv:0907.1919}}.

\bibitem{Kazakov:2009stock}
V.~Kazakov and N.~Gromov, ``{Talk at IGST-2010, Nordita, Stockholm}'', {\em
  {http://agenda.albanova.se/contributionDisplay.py?contribId=258\&confId=1561%
}}  .

\bibitem{Frolov:2010talk}
S.~Frolov, ``{Talk at QFT2010, DESY, Hamburg}'', {\em
  {http://qft2010.desy.de/sites/site\_qft2010/content/e88901/e104143/infoboxCo%
ntent104215/Frolov\_DESY10.pdf}}  .

\bibitem{Beccaria:2008tg}
M.~Beccaria, V.~Forini, A.~Tirziu, and A.~A. Tseytlin, ``{Structure of large
  spin expansion of anomalous dimensions at strong coupling}'',
  \href{http://dx.doi.org/10.1016/j.nuclphysb.2008.12.013}{{\em Nucl. Phys.}
  {\bf B812} (2009)  144--180},
\href{http://arXiv.org/abs/0809.5234}{\texttt{arXiv:0809.5234}}.

\bibitem{Fioravanti:2010qh}
D.~Fioravanti, P.~Grinza, and M.~Rossi, ``{Beyond cusp anomalous dimension from
  integrability in SYM$_4$}'',
\href{http://arXiv.org/abs/1011.4005}{\texttt{arXiv:1011.4005}}.

\bibitem{Korchemsky:1995be}
G.~P. Korchemsky, ``{Quasiclassical QCD pomeron}'',
  \href{http://dx.doi.org/10.1016/0550-3213(96)00019-3}{{\em Nucl. Phys.} {\bf
  B462} (1996)  333--388},
\href{http://arXiv.org/abs/hep-th/9508025}{\texttt{arXiv:hep-th/9508025}}.

\bibitem{Roiban:2011fe}
R.~Roiban and A.~A. Tseytlin, ``{Semiclassical string computation of
  strong-coupling corrections to dimensions of operators in Konishi
  multiplet}'',
\href{http://arXiv.org/abs/1102.1209}{\texttt{arXiv:1102.1209}}.

\bibitem{Vallilo:2011fj}
B.~C. Vallilo and L.~Mazzucato, ``{The Konishi multiplet at strong coupling}'',
\href{http://arXiv.org/abs/1102.1219}{\texttt{arXiv:1102.1219}}.

\end{thebibliography}\endgroup

\end{document}